\documentclass[preprint,12pt]{elsarticle}

\usepackage{amssymb}
\usepackage{amsmath}
\usepackage{xcolor}

\usepackage{booktabs}
\usepackage{float}
\usepackage{hyperref} 
\usepackage{pdfpages}
\usepackage{subcaption}
\usepackage{multirow}
\usepackage{xcolor}
\usepackage{float} 

\usepackage{natbib}

\usepackage{natbib}  
\usepackage{url}

\begin{document}
\begin{frontmatter}

\title{Diagnostically Competitive Performance of a Physiology-Informed Generative Multi-Task Network for Contrast-Free CT Perfusion}

\author[1]{Wasif Khan}
\author[3] {John Rees}
\author[1]{Kyle B. See}
\author[10]{Simon Kato}
\author[1]{Ziqian Huang}
\author[1]{Amy Lazarte}
\author[1]{Kyle Douglas}
\author[2] {Xiangyang Lou}
\author[8] {Teng J. Peng}
\author[3] {Dhanashree Rajderkar}
\author[4,5,6] {Pina Sanelli}
\author[8] {Amita Singh}
\author[3] {Ibrahim Tuna}
\author[8] {Christina A. Wilson}

\author[1,7,9]{Ruogu Fang\corref{Ruogu Fang}}


\affiliation[1]{
    organization={J. Crayton Pruitt Family Department of Biomedical Engineering, University of Florida},
    city={Gainesville},
    state={FL},
    country={USA}
}
\affiliation[3]{
    organization={Department of Radiology, University of Florida},
    city={Gainesville},
    state={FL},
    country={USA}
}
\affiliation[10]{
    organization={Department of Mathematics and Statistics, University of Florida},
    city={Gainesville},
    state={FL},
    country={USA}
}

\affiliation[2]{
    organization={Department of Biostatistics, University of Florida},
    city={Gainesville},
    state={FL},
    postcode={32611},
    country={USA}
}
\affiliation[8]{
    organization={Department of Neurology, University of Florida},
    city={Gainesville},
    state={FL},
    country={USA}
}

\affiliation[4]{
    organization={Feinstein Institutes for Medical Research},
    city={Manhasset},
    state={NY},
    country={USA}
}
\affiliation[5]{
    organization={Department of Radiology, Donald and Barbara Zucker School of Medicine at Hofstra Northwell},
    city={Hempstead},
    state={NY},
    country={USA}
}
\affiliation[6]{
    organization={Northwell Health},
    city={Manhasset},
    state={NY},
    country={USA}
}
\affiliation[7]{
    organization={Center for Cognitive Aging and Memory, McKnight Brain Institute, University of Florida},
    city={Gainesville},
    state={FL},
    country={USA}
}
\affiliation[9]{
    organization={Department of Electrical and Computer Engineering, University of Florida},
    city={Gainesville},
    state={FL},
    country={USA}
}



\cortext[Ruogu Fang]{Corresponding author}
\ead{ruogu.fang@bme.ufl.edu}

\begin{abstract}
Perfusion imaging is extensively utilized to assess hemodynamic status and tissue perfusion in various organs. Computed tomography perfusion (CTP) imaging plays a key role in the early assessment and planning of stroke treatment. While CTP provides essential perfusion parameters to identify abnormal blood flow in the brain. However, CTP can be expensive with limited accessibility, and the use of contrast agents in CTP can lead to allergic reactions and adverse side effects. To address these challenges, we propose a novel deep learning framework called Multitask Automated Generation of Intermodal CT perfusion maps (MAGIC). This framework combines generative artificial intelligence and physiological information to map non-contrast computed tomography (CT) imaging to multiple contrast-free CTP imaging maps. We demonstrate enhanced image fidelity by incorporating physiological characteristics into the loss terms. Our network was trained and validated using CT image data from patients referred for stroke at UF Health and demonstrated robustness to abnormalities in brain perfusion activity. A double-blinded study was conducted involving seven experienced neuroradiologists and vascular neurologists. This study validated MAGIC’s visual quality and diagnostic accuracy showing favorable performance compared to clinical perfusion imaging with intravenous contrast injection. Overall, MAGIC holds great promise in revolutionizing healthcare by offering contrast-free, cost-effective, and rapid perfusion imaging.
\end{abstract}


\end{frontmatter}


\section{Introduction}

Generative artificial intelligence (AI) has emerged as a highly promising tool in recent years. This has been exemplified by groundbreaking technologies such as ChatGPT and DALL-E. Its potential for advancing healthcare is immense, ranging from medical image synthesis and restoration to segmentation and object transfiguration \cite{yang_intelligent_2021,doi_computer-aided_2007,sudarshan_low-dose_2021,wang_deep_2021,shan_competitive_2019}. In this study, we demonstrate the potential feasibility of generative AI in enhancing the efficiency of clinical diagnostic workflows and reducing potential risks from contrast agent injection. The proposed generative AI-based contrast-free perfusion image generation is a complex task that involves transferring an image from a non-contrast image domain to a contrast-enhanced image domain, while preserving essential clinical representations \cite{zhu_unpaired_2017}. 

Medical image-to-image translation, particularly in synthetic contrast enhancement, has been a focus of research in recent years. For instance, studies have demonstrated the successful use of deep learning techniques to generate synthetic contrast-enhanced images from non-contrast scans, such as in cardiac CT \cite{santini2018synthetic} and MRI \cite{denck2021mr} modalities. These techniques aim to improve clinical workflow by reducing the need for contrast agents, thereby minimizing the associated risks. This approach has been explored using various deep learning architectures, including convolutional neural networks (CNNs) and generative adversarial networks (GANs), to synthesize high-quality images for clinical use \cite{rofena2024deep, chen2022synthesizing, singh2021medical, fard2022cnns, armanious2020medgan, kazeminia2020gans, hognon2024contrastive}. However, none of them use non-contrast imaging to generate the hemodynamic perfusion maps using deep learning. Furthermore, many of these methods often struggle with maintaining clinical accuracy and detail when translating between image domains, particularly in perfusion imaging, where fine-grained vascular information is crucial for diagnosis. Therefore, we present a novel perspective that holds the potential to revolutionize the approach to diagnosing and treating blood flow and vascular conditions.

Computed tomography perfusion (CTP) is a potent imaging modality widely used to evaluate hemodynamic properties and to detect abnormalities in the brain parenchyma, particularly in the context of acute ischemic stroke \cite{konstas_theoretic_2009}. CTP can provide valuable information about blood flow and blood volume to various vascular territories and can indicate hypo-perfusion. Hypo-perfusion can be seen in ischemic stroke when a cerebral artery becomes occluded \cite{khandelwal_ct_2008,nguyen_noncontrast_2022}. In addition, perfusion imaging has widespread utility across other organs including: the liver, kidney, prostate, pancreas, soft tissue, spleen, heart, and bone \cite{ogul_perfusion_2014,yuan_simplified_2016,somford_diffusion_2008,perik_quantitative_2022,emanuel_contrastenhanced_2020, vancauwenberghe_imaging_2015,nieman_dynamic_2020,lee_assessment_2009,ogul_abdominal_2013}. Perfusion imaging allows for the identification of abnormal blood flow and volume to aid in the diagnosis, prognosis, and management of cardiovascular, renal, oncological, and many other disorders \cite{nieman_dynamic_2020,zhang_renal_2020,petralia_ct_2010}.

Stroke is a second leading cause of death in the world \cite{noauthor_top_nodate} with more than 795,500 strokes occur in the U.S. each year \cite{noauthor_stroke_2023}. Prompt treatment of stroke patients has been associated with meaningful improved outcomes, allowing patients to return to their families and resume normal life. Therefore, identifying and characterizing the hypo-perfused tissue in acute ischemic stroke is crucial to ensuring the best clinical outcomes \cite{saver2006time}. Non-contrast computed tomography (NCCT) is used to exclude diagnoses of brain tumors and intracerebral hemorrhages but other advanced modalities such as CT angiography and CTP are necessary to accurately identify patients with acute ischemic stroke \cite{vagal2014increasing}. Up to 40\% of stroke patients have a normal NCCT scan within the first few hours \cite{noauthor_early_nodate} of stroke, highlighting the need of CTP in assessing cerebral perfusion and its value in identifying potentially salvageable brain tissue \cite{bill2019focal}. CTP imaging works by imaging a volume of brain tissue during the intravenous injection of an iodinated contrast bolus. These data are used to characterize the capillary-level perfusion within the brain and identify any regions of abnormal blood flow and blood volume. Thus, CTP is important for evaluating the stroke status and informing therapeutic decision making. Prompt therapy can then be used to restore blood flow to the tissue through the use of tissue plasminogen activator or thrombectomy \cite{khandelwal_ct_2008,federau_ct_nodate}. Additionally, the mismatch between cerebral blood flow / cerebral blood volume or mean transit time / cerebral blood volume can be used to identify salvageable brain tissue that may recover. However, even though CTP is beneficial, this imaging procedure is lengthy, requires contrast agent injection, and may delay stroke treatment. Moreover, many healthcare settings—especially rural or low-resource areas—do not have access to perfusion-capable CT scanners, and not all CT scanners support CTP protocols \cite{bergeron2017lack}. Mobile stroke units, which are becoming increasingly common, also typically rely on standard NCCT rather than full CTP due to equipment and time constraints. This lack of access can hinder timely decision-making for interventions such as thrombectomy, especially when preparing for the operating room where every minute counts. Having CT perfusion maps generated by AI methods could assist clinicians to accelerate treatment and improve outcomes. Therefore, we propose the use of generative artificial intelligence and physiological information to map NCCT imaging to contrast-free CTP imaging maps.


\section{Methods}
\subsection{Data}
\subsubsection{Data acquisition}
Neuroimaging data were retrospectively retrieved at UF Health between 2013 and 2022 with Institutional Review Board (IRB) approval by the University of Florida. Written informed consent was waived by the IRB. The perfusion studies were conducted using an injection of an iodine-based contrast agent, and the collected data were analyzed using RAPID CTP analysis software (iSchemaView, Menlo Park, CA., USA) to generate each CTP map. The data were queried to include patients whose imaging data included RAPID CTP data. A total of 880 patients were found to have both RAPID CTP and NCCT imaging data. After excluding 39 patients due to incomplete data, 841 patients remained for analysis.

Among them, 673 patients (80\%) were randomly selected for training. The remaining 168 patients (20\%) were reserved as an independent test set; and 20 patients from the test set were further randomly selected for diagnostic evaluation. Furthermore, static  data augmentation was applied to the training data \cite{hao2021comprehensive}: random horizontal reflection, random rotation between -10 degrees and 10 degrees, and random vertical and horizontal translations between -30 and 30 pixels in each direction \cite{chen_what_2021}. After extracting and applying augmentations to paired NCCT and CTP data, 29,358 (26.8 GB) paired slices of NCCT and CTP images were used to train MAGIC. Each image was normalized with a mean of 0 and a standard deviation of 1 in the training environment.

\subsubsection{Preprocessing}
The axial direction spatial resolution is 10 mm for CTP volumes and 1 mm for NCCT volumes. 10 CTP slices located at central axial locations were selected for analysis for each patient. All slices in the CTP series were coordinated with one NCCT slice at the corresponding vertical location and two NCCT slices at a predefined vertical offset of 4mm between slices. The three NCCT slices were stacked to create a pseudo-RGB image. MAGIC can accurately synthesize higher quality cerebral perfusion structures , largely due to the additional spatial information above and below the target slice. Additionally, each patient’s NCCT series was collected. The data were deidentified, and all images were resized to [256×256] pixels. We normalized the scales for each perfusion map type using the following physiological ranges to map perfusion map values to normalized pixel values: CBF: [0, 60] (mL/100 g/min); CBV: [0, 4] (4 mL/100 g); MTT: [0, 12] (s); TTP: [0, 25] (s). For the NCCT series, each slice was rescaled with a window center of 40 HU and a window width of 80 HU. Skull-stripping was conducted on each NCCT slice to focus on the brain tissues for further analysis. 

\subsubsection{Ischemia severity}
The patients in this dataset presented a variety of stroke presentations, ranging from healthy to severely infarcted. RAPID yields diagrams for each patient of identified regions of infarct core (defined by CBF \(<\) 30\%) and ischemic penumbra (defined by Tmax \(>\) 6s)\cite{rapid_nodate}. Studies have shown that these are accurate surrogate regions for the infarct core and ischemic penumbra regions, which we use as surrogates for the infarct core and ischemic penumbra regions below. The indicated regions in the RAPID data were used to create ROI masks to further evaluate MAGIC’s performance in encoding infarcted brain regions. To evaluate the infarction severity of each subject, the ratio of infarct core and ischemic penumbra was defined as the ratio of infarct core to ischemic penumbra in the 3-dimensional volume of the whole brain. This ratio is also called RAPID Ratio or the Mismatch Ratio (MMR) \cite{demeestere_review_2020,vagal_automated_2019,chen_what_2021,potter_ct_2019}.  MMR was calculated by summing the number of pixels of reduced CBF and delayed TTP volumes respectively for each axial slices in the CBF/Mismatch analysis. These ratios were used to classify each patient as either presenting no infarction, or with different levels of infarction. The class of no infarction represents core-to-penumbra ratio of 0, and with infarction group has a core-to-penumbra ratio greater than 0. Among the infarction group, mild to moderate infarction is defined by a core-to-penumbra ratio larger than 0 but less than or equal to 0.5, and severe group has a ratio of greater than 0.5. These ratio classes were determined with the inspiration of Straka et al.\cite{straka_real-time_2010} and exactly chosen with the help of the radiologists.

\subsection{MAGIC}
In this study, we introduce MAGIC, a model designed to generate individual perfusion parametric maps – CBV, CBF, MTT, and TTP – from NCCT. Our approach aims to improve the efficiency, accessibility, safety, and cost cost-effectiveness of CTP imaging.

\subsubsection{Perfusion maps}
Here we introduce the definition of each perfusion map briefly. CBV is a measure of blood volume in a brain tissue region and is typically measured in mL of blood per 100g of brain tissue \cite{khandelwal_ct_2008}. CBF is a measure of the rate at which blood travels through the brain tissue and is typically measured in units of mL of blood per 100g of tissue per minute \cite{kingma_adam_2017}.  MTT is a measure of the mean bolus transit time through a volume of brain tissue from the arterial input to the venous output and is typically measured in seconds \cite{khandelwal_ct_2008}. TTP is a measure of the time required for a given tissue volume to reach maximal enhancement and is typically measured in seconds \cite{khandelwal_ct_2008,lin_perfusion_2013}. CTP imaging is valuable in assessing diseases characterized by altered brain perfusion activity, including ischemic stroke. The core infarct of ischemic stroke is characterized by increased MTT and TTP and decreased CBF and CBV \cite{lin_perfusion_2013}. The penumbra of ischemic stroke is characterized by increased MTT and TTP with sufficient CBF and CBV. \cite{lin_perfusion_2013}.  A bounding box was applied to the brain region in both the synthesized and real perfusion imaging to ensure that results would not be skewed by the background space surrounding the brain tissue in each image.

\begin{figure}[H]
\centering
\includegraphics[trim=12cm 20cm 22cm 3cm, clip,width=0.9\linewidth]{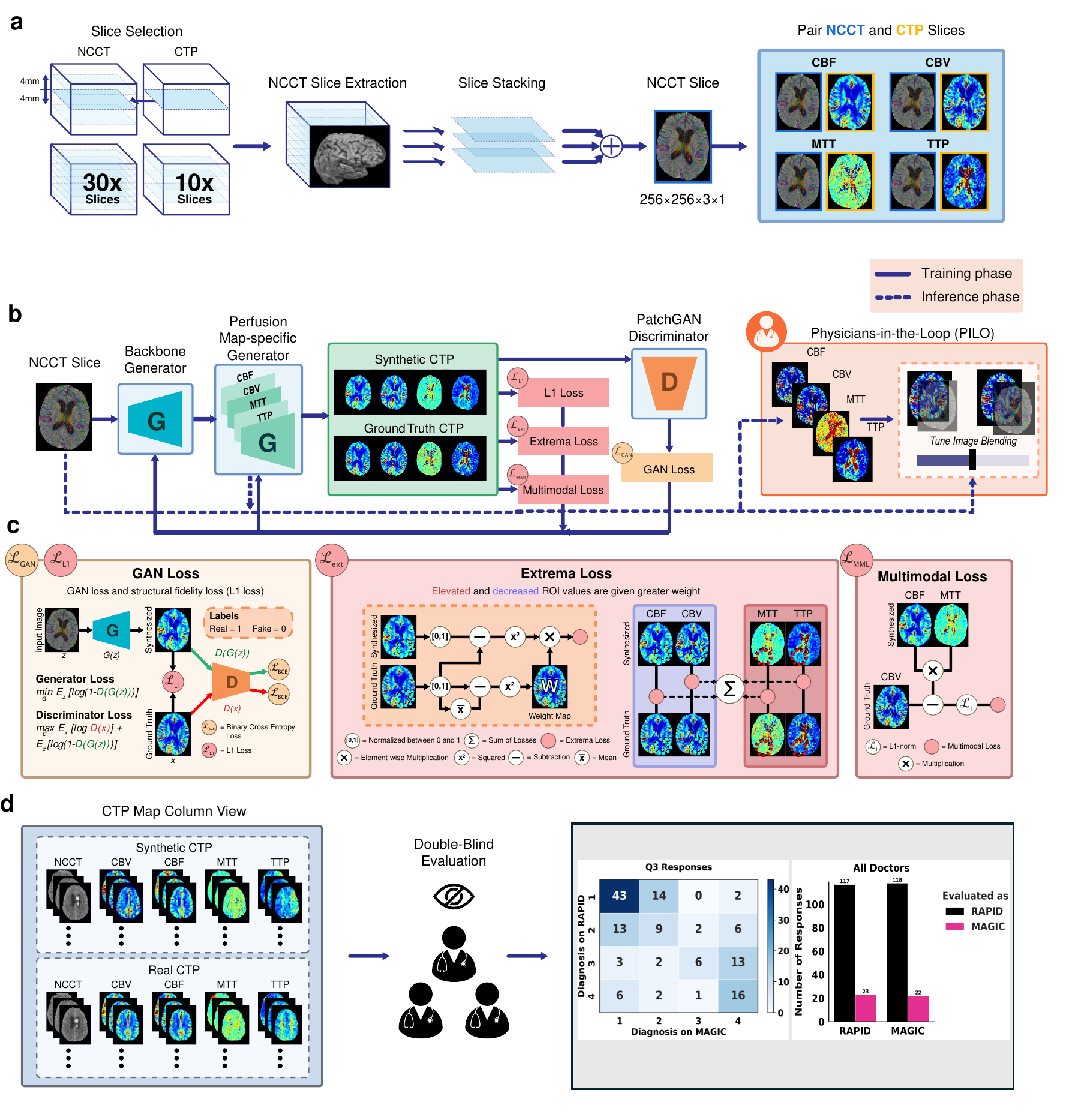}
\caption{Proposed MAGIC pipeline for non-contrast CT perfusion map generation. \textbf{a}. Neighboring NCCT slices in the z-axis are stacked to create a pseudo-RGB image and are matched with the corresponding perfusion maps to construct paired datasets. \textbf{b}. MAGIC consists of a generative adversarial network to co-learn between the generator and discriminator networks using novel physiology-informed loss terms. During training, the generator synthesizes perfusion maps using a UNet architecture, which are refined through adversarial learning with the PatchGAN discriminator. In the testing stage — indicated by the dashed line in the figure —the Physician-in-the-Loop (PILO) interactive module allows clinicians to visually assess and adjust the optimal ratio between non-contrast structural information and synthesized perfusion information.} \textbf{c}. The composite loss terms contribute synergistically to MAGIC's high-quality performance. During training, GAN loss (L\textsubscript{GAN}) and L1 loss (L\textsubscript{L1}) are used to ensure style consistency and structural fidelity in the synthesized imaging. The novel extrema loss (L\textsubscript{ext}) is used to ensure proper encoding of ischemic tissue regions. The novel multimodal loss (L\textsubscript{MML}) is used to ensure physiological consistency across the synthesized perfusion maps. \textbf{d}. It was determined that seven experienced neuroradiologists could not differentiate the synthetic and real CT perfusion imaging in a double-blinded review in terms of the realistic level and diagnostic quality.
\label{fig:overview}
\end{figure}

A novel GAN architecture is proposed for the task of NCCT to CTP translation. MAGIC is based on the traditional pix2pix architecture with novel multitask architecture and physiology-informed loss functions as an image-to-image translation network \cite{isola_image--image_2018}. In contrast to traditional image translation networks, MAGIC enables greater performance in this task due to 1) simultaneous generation of multiple CTP maps, 2) shared encoding layers across all perfusion maps, allowing the generator to extract common anatomical features and reduce the number of trainable parameters, 3) physiology-informed loss terms that leverage MAGIC’s multitask architecture, 4) real-time generation of perfusion maps without contrast agent injection, 5) high spatial resolution in the axial direction and 3D models, 6) generalizability to any additional perfusion maps, and 7) a novel Physician-In-The-Loop (PILO) module, which is a tunable convolution layer to provide a physician-computer interface to allow physicians adjust the contribution of perfusion and anatomical information in the synthesized CTP maps to facilitate clinical decision making. The overview of the proposed method is presented in Fig. \ref{fig:overview}.

\subsubsection{Generator}
The generator of MAGIC is inspired by a U-NET convolutional network architecture, which consists of a series of convolutional (encoding) layers that progressively capture hierarchical features and transposed convolution (decoding) layers that reconstruct the spatial structure of the input. These encoding and decoding layers are paired using skip connections, which directly transfer spatial information from the encoder to the corresponding decoder layers, helping to retain fine-grained details and improve reconstruction quality \cite{ronneberger_u-net_2015}. The generator architecture enables MAGIC’s multitask learning by applying a series of four encoding layers to a given NCCT input (Supplementary Fig. S.1). Following these shared encoding layers, four distinct encoding layers are applied to produce four final encoded tensors, each with a size of [1x1x512]. Each tensor is passed through a unique series of eight decoding layers to produce an output tensor of size [256×256×1]. Finally, each output tensor is concatenated with the original NCCT input of size [256×256×3] to produce a [256×256×4] tensor, which is passed through the PILO module. This module reduces the dimensions of the final tensor to [256×256×1], producing the corresponding CTP map slice. 


\subsubsection{Discriminator}
The discriminator of MAGIC is based on PatchGAN, a widely used CNN-based architecture designed for image-to-image translation tasks \cite{isola_image--image_2018,perik_quantitative_2022}. 
Unlike conventional discriminators that classify entire images as real or synthesized, PatchGAN operates at the patch level, evaluating local image regions to determine authenticity. This approach encourages the generator to produce more locally consistent textures and fine-grained details, making it effective for medical image synthesis. For this task, PatchGAN was used with a receptive field size of [70×70] to sufficiently evaluate the “realistic” quality of the synthesized images. The input tensor for PatchGAN is a [256×256×4] tensor containing the input image concatenated with its corresponding label. The input for a synthesized image is concatenated with a matrix of zeros along the first dimension. Similarly, the input is concatenated with a matrix of ones along the first dimension for a real image. 
PatchGAN outputs a patch-wise probability map, where each value represents the likelihood of a corresponding image patch being real. These outputs are used to compute the adversarial loss, which guides both the generator and discriminator training in MAGIC. The architecture of the discriminator network is illustrated in Supplementary Fig. S.2.


\subsubsection{Physiology-Informed Loss Functions}
GAN training typically represents a zero-sum game in which a generator network and a discriminator network are updated dynamically to learn the mapping between multiple image domains. The generator learns to synthesize accurate CTP maps during this training process given an NCCT input. The discriminator learns to distinguish the generator’s output from the real data distribution. Additionally, we propose two novel loss terms: a multimodal loss and an extrema loss. The multimodal loss term is inspired by the physiological relationship between CTP map classes. The extrema loss term encodes regions of tissue ischemia. Both work together to increase the accuracy of the synthesized maps.

\subsection{Loss Functions}
\subsubsection{Notations}
We adopt a similar notation to Zhu et al \cite{zhu_unpaired_2017}. The goal of our modal is to learn four mapping functions between the domain X of NCCT images and the domain of perfusion maps (CBV, CBF, MTT, and TTP). Given M training samples, each is composed of five elements: \({\{x_i,cbv_i,cbf_i,mtt_i,ttp_i \}}_i^M\) for the \(i^{th}\) sample. Here \(x_i \in X\) where X represents the set of NCCT images, \(cbv_i \in CBV\) , \(cbf_i \in CBF\), \(mtt_l \in MTT\), and \(ttp_m \in TTP\). We will refer to the set containing all four perfusion maps as \(CTP={CBV,CBF,MTT,TTP}\). Let N denote the cardinality of CTP which is 4 in our case but can be flexible depending on how many perfusion maps to predict for the generalizability of this method. Recalling that X contains the real NCCT, we denote the input data distribution as \(x~p_{data} (x)\) and will indicate the data distribution of the different perfusion maps as \(cbv \sim p_{data} (cbv)\),\(cbf \sim p_{data} (cbf)\),\(mtt \sim p_{data} (mtt)\), and \(ttp \sim p_{data} (ttp)\). Importantly, we denote the joint data distribution of an NCCT and its corresponding real perfusion map as \((x,map) \sim (p_{data} (x),p_{data} (map))\) where map can be replaced by any type from \({cbv, cbf, mtt, ttp}\) depending on the perfusion map being discussed. The model includes four mappings represented by four generators: \(G_{CBV}:X \rightarrow CBV\), \(G_{CBF}:X \rightarrow CBF\), \(G_{MTT}:X \rightarrow MTT\), and \(G_{TTP}:X \rightarrow TTP\). Moreover, the model has four discriminators \(D_{CBV}\),\(D_{CBF}\),\(D_{MTT}\),and \(D_{TTP}\) where each discriminator aims to distinguish between its corresponding real perfusion map and the synthetic perfusion map. Finally, we use \(\odot\) to represent element-wise multiplication and \(W_x\) to represent a weighted matrix of a perfusion map of a given input x (real NCCT image). The weight matrices are constructed by taking a real perfusion map image, normalizing the pixel values, and squaring so all pixel values are positive. 

\subsubsection{Physiology-Informed Loss Functions}
Traditional generators in GANs utilize a sum of loss terms to evaluate 1) the similarity between the synthesized and the real images and 2) the generator’s ability to “fool” the discriminator. We use a modified form of the objective function proposed by Isola et al \cite{isola_image--image_2018}. that mixes the traditional GAN objective with a structural fidelity loss term. The GAN term, as provided in Eq. (\ref{eq:GAN}), enables the adversarial nature of the MAGIC network, as \(G_{CTP}\) aims to minimize the objective while \(D_{CTP}\) aims to maximize it. \(D_{CTP}\) is incentivized to correctly classify synthetic and real CTP maps, and \(G_{CTP}\) is incentivized to produce outputs that fool the discriminator into incorrectly classifying synthetic CTP maps as real CTP maps. As an implementation detail, the weights of \(D_{CTP}\) are first optimized and frozen before optimizing the weights of \(G_{CTP}\). The structural fidelity loss term, as provided in Eq. (\ref{eq:L1}), is defined as the L1 distance between the generated map and the real map. While the inclusion of this term does not affect the task of the discriminator, it ensures that the generator produces outputs that minimize the L1 distance between the generated and real CTP maps.


\begin{flalign}
\label{eq:GAN}
&  \begin{split} 
\mathcal{L}_{\text{GAN}} (G_{CTP}, D_{CTP}, CTP, X) = & \frac{1}{N} \sum_{PM \in CTP} \mathbb{E}_{PM \sim p_{\text{data}}(PM)} \left[ \log D_{PM}(PM) \right] \\
& + \mathbb{E}_{x \sim p_{\text{data}}(x)} \left[ \log(1 - D_{PM}(G_{PM}(x))) \right]
\end{split} &
\end{flalign}


\begin{flalign}
\label{eq:L1}
& \begin{split}
\mathcal{L}_{L1}(G_{CTP}, CTP, X) = & \frac{1}{N} \sum_{PM \in CTP} \mathbb{E}_{(x, PM) \sim (p_{\text{data}}(x), p_{\text{data}}(PM))} \left[ \| PM - G_{PM}(x) \|_{1} \right]
\end{split} &
\end{flalign}

We propose two novel loss terms, multimodal loss and extrema loss, in MAGIC to leverage the physiological rules between CTP map types to further improve performance. Specifically, the multimodal loss leverages the Central Volume Principle \cite{hoeffner_cerebral_2004}, which describes the physiological relationship between the CBF, CBV, and MTT map classes. These map types are related by Eq. (\ref{eq:relation}), which leads to the multimodal loss represented in Eq. (\ref{eq:MML}). The generator minimizes the L1 distance between the real CBV map and the product of the synthesized CBF and MTT maps, ensuring that the synthetic CTP maps are consistent with the central volume principle. 


\begin{flalign}
&CBV = CBF \times MTT && \label{eq:relation}
\end{flalign}


\begin{flalign}
\label{eq:MML_final_split}
& \begin{split}
\mathcal{L}_{\text{MML}}(G_{CBF}, G_{MTT}, CBV, X) = & \mathbb{E}_{(x, CBV) \sim (p_{\text{data}}(x), p_{\text{data}}(CBV))} \\
& \quad \left[ \| G_{CBF}(x) \times G_{MTT}(x) - CBV \|_{1} \right]
\end{split} &
\end{flalign}

The extrema loss term enables increased fidelity in encoding regions-of-interest (ROI) in the simulated scans that deviate from the mean of the observed image. The extrema loss was introduced to ensure that MAGIC would accurately generate the ROIs with both very high and very low pixel intensities compared to the slice mean value. Its inclusion increases the likelihood that stroke regions in the synthesized perfusion imaging are properly identified and learned by MAGIC during training. Infarcted tissue regions tend to exhibit elevated MTT and TTP and decreased CBF and CBV, so these characteristics must be present in the synthetic CTP maps to ensure acceptable diagnostic quality. We generated a weight map \(W_x\) by normalizing the real perfusion map between [-0.5, 0.5] and then take an element-wise square of the normalized image, as shown in Eq. (\ref{eq:wx}). \(W_x\) emphasizes the ROIs with values farther from the image mean. Then this weight map is multiplied by the MSE between the real and synthesized perfusion imaging, shown in Eq. (\ref{eq:hx}). The mean of this product is taken to produce the final extrema loss term, as provided in Eq. (\ref{eq:EXT}). The weight matrices are constructed by taking a real perfusion map image, normalizing the pixel values, and squaring so all pixel values are positive.



\begin{flalign}
& W_{x} = \left( \frac{PM - \min(PM)}{\max(PM) - \min(PM)} - 0.5 \right)^{2} && \label{eq:wx_left}
\end{flalign}


\begin{flalign}
& H_{x} = \left(\frac{G_{PM}(x) - \min(G_{PM}(x))}{\max(G_{PM}(x)) - \min(G_{PM}(x))} - \frac{PM - \min(PM)}{\max(PM) - \min(PM)}\right)^{2} && \label{eq:hx}
\end{flalign}

\begin{equation}
\mathcal{L}_{EXT}(G_{CTP},CTP,X)=\frac{1}{N}\sum_{PM\in CTP}\mathbb{E}_{(x,P M)\sim(p_{data}(x),p_{data}(PM))}[W_{x}\odot H_{x}]
\label{eq:EXT}
\end{equation}

We weight each additional loss term with lambda tuning hyperparameters. The final objective we aim to solve, which \(G_{CTP}\) aims to minimize and \(D_{CTP}\) aims to maximize, is shown below in Eq. (\ref{eq:g*}).


\begin{flalign}
\label{eq:g*}
& G^{\ast} = \arg \min_{G_{CTP}} \max_{D_{CTP}} \mathcal{L}_{GAN}(G_{CTP},D_{CTP},CTP,X) & \\
&   \quad + \lambda_{1}\mathcal{L}_{L1}(G_{CTP},CTP,X) \notag \\
&   \quad + \lambda_{2}\mathcal{L}_{EXT}(G_{CTP},CTP,X) \notag \\
&   \quad + \lambda_{3}\mathcal{L}_{MML}(G_{CBF},G_{MTT},CBV,X) \notag
\end{flalign}

\subsubsection{Training Process}
The training of MAGIC follows a standard GAN framework. For each iteration, the multitask generator G synthesizes CTP maps from a given NCCT input, the discriminator D evaluates both real (RAPID) and synthetic (MAGIC) pair tensors. The discriminator and generator losses are computed, and their respective weights are updated. This process is repeated until convergence of both losses. During inference, the perfusion information ratio in the PILO module can be scaled to select the optimal contribution from anatomical and hemodynamic imaging.

\subsubsection{Network Training}

Following the design of the original pix2pix framework \cite{isola_image--image_2018}, each encoding layer employs a $4\times 4$ convolutional kernel. A stride of 2 and zero-padding of 1 are applied to maintain spatial dimensions during convolution. The convolutional layers in the generator use LeakyReLU activations, and Tanh activation in the final layer. In the discriminator, all intermediate convolutional layers use LeakyReLU, while the final convolution outputs a probability map through a Sigmoid activation. Batch Normalization was utilized when the batch size was greater than 1, and Instance Normalization was used when the batch size was 1. The generator has 209,352,408 total parameters while the discriminator has 11,063,044 total parameters. MAGIC was implemented using the PyTorch version 1.7.0 library in Python version 3.7.6 and the NVIDIA CUDA platform version 11.7.1 to enable GPU-accelerated computation. MAGIC was trained for 50 epochs on an NVIDIA DGX A100 with a GPU RAM of 80 GB. An Adam optimizer with \(\beta\) values of \(\beta 1=0.5\) and \(\beta2=0.999\) were used \cite{newcombe_interval_1998}. After 50 epochs of training, performance was not found to improve further. A batch size of 8 was used. The learning rates for the generator and discriminator were set to a constant value of \(1e^{-4}\). A learning rate decay was not found to improve the quality of results. After each epoch, the generator loss was computed and recorded on a validation set of 100 randomly selected sample images (at the image level) to ensure that the MAGIC was not overfitting the training set. In total, model training spanned around 40 hours.

\subsection{Evaluation Metrics}
The evaluation of MAGIC’s performance consists of two primary components to assess the overall quality and clinical applicability of the synthesized CTP series: 1) quantitative image evaluation and 2) expert visual assessment. 

\subsubsection{SSIM and UQI}
The primary image quality metrics used to define MAGIC’s performance in synthesizing high-quality CTP series are structural similarity index metric (SSIM) and universal quality index (UQI) \cite{nilsson_understanding_2020,wang_universal_2002}. SSIM and UQI are preferred metrics for assessing the clinical viability of the synthetic perfusion imaging results as they provide a full-real measure of perceptual image quality based on factors like local structure, contrast, and luminance \cite{nilsson_understanding_2020,wang_universal_2002}. Unlike metrics that rely on absolute pixel-wise differences, such as PSNR, SSIM and UQI are less sensitive to global intensity variations caused by factors like slight patient movements between NCCT and CTP series acquisition. Metrics like PSNR, which rely on pixel-wise intensity comparisons, may not accurately reflect the perceptual similarity or diagnostic utility of the generated images. SSIM and UQI, in contrast,   offer a more clinically relevant evaluation by focusing on structural integrity and perceptual similarity rather than absolute intensity differences. A higher perceptual similarity score between synthetic and real perfusion imaging suggests that the synthetic images could yield comparable diagnostic outcomes when assessed by clinicians \cite{renieblas_structural_2017}. SSIM and UQI are defined in Eq. (\ref{eq:SSIM}) and (\ref{eq:UQI}), respectively. In Equation~(\ref{eq:SSIM}), the parameters $c_1$ and $c_2$ are small stabilizing constants used to avoid division by zero when the image means or variances are very small. Specifically, $c_1$ stabilizes the luminance term $(\mu_x^2 + \mu_y^2)$ and $c_2$ stabilizes the contrast/structure term $(\sigma_x^2 + \sigma_y^2)$ \cite{wang_universal_2002}.


\begin{flalign}
& SSIM(x,y)=\frac{(2\mu_{x}\mu_{y}+c_{1})(2\sigma_{xy}+c_{2})}{(\mu_{x}{}^{2}+\mu_{y}{}^{2}+c_{1})(\sigma_{x}{}^{2}+\sigma_{y}{}^{2}+c_{2})} && \label{eq:SSIM}
\end{flalign}

\begin{flalign}
& UQI(x,y)=\frac{4\sigma_{xy}\bar{x}\ \bar{y}}{(\sigma_{x}{}^{2}+\sigma_{y}{}^{2})[(\bar{x})^{2}+(\bar{y})^{2}]}
 &&\label{eq:UQI}
\end{flalign}


\subsubsection{Double-blind Clinical Evaluation}
\begin{figure}[!htbp]
\centering
\includegraphics[trim=4cm 2cm 6cm 3cm, clip,width=\linewidth]{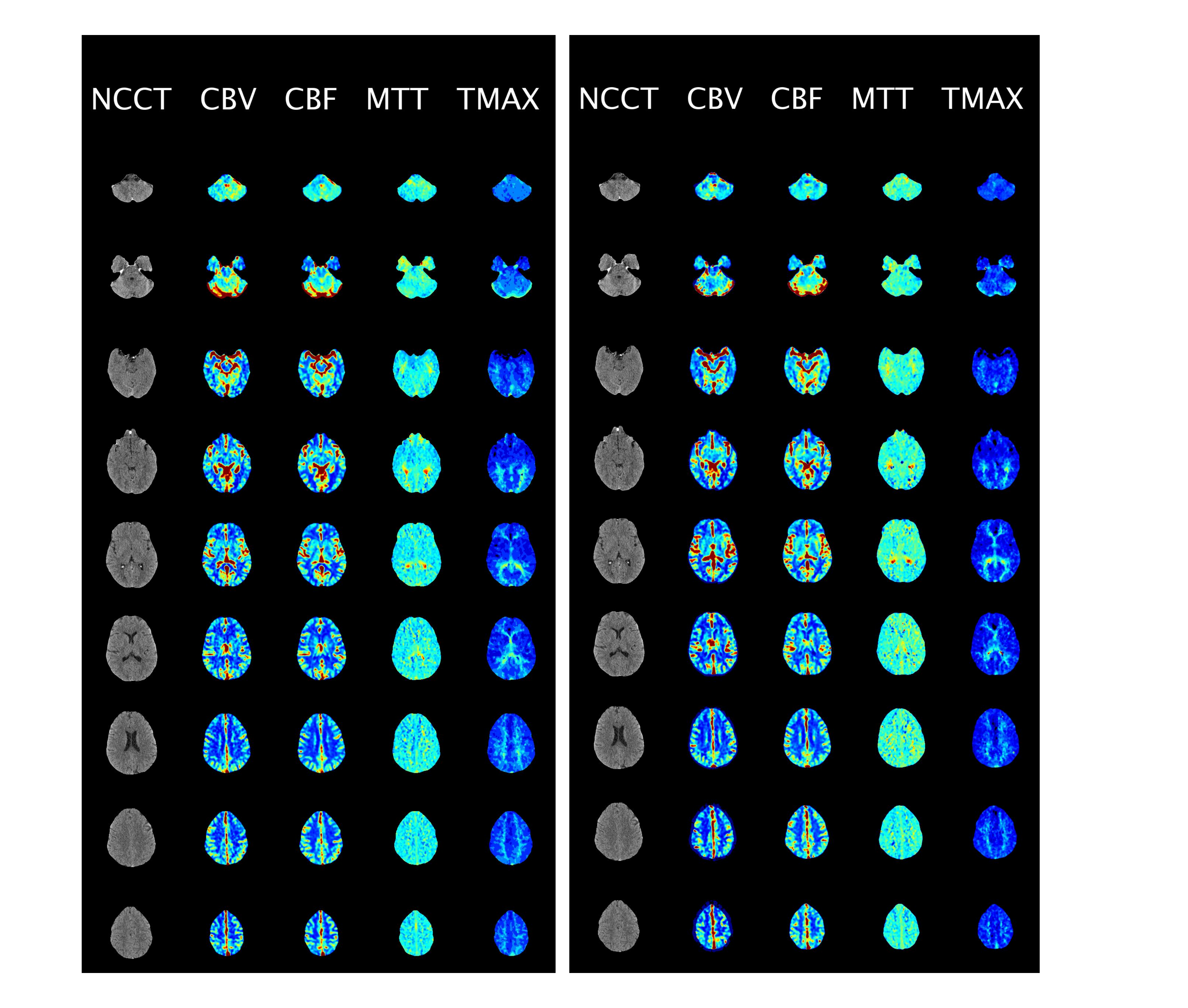}
\caption{Sample CTP maps in doubled blinded evaluations. The CTP perfusion maps were presented in column view and given to clinicians for clinical evaluation of a patient with moderate infarction. The left image is the real (RAPID) CTP maps, and the right image is the synthetic (MAGIC) CTP maps for the same patient.}
\label{fig:double_blind_example}
\end{figure}

Twenty patients from the test data were used to conduct the experiments. These stratifications were determined using region of interest (ROI) data from RAPID. The experiment was split into two trials to prevent bias due to reviewing both the RAPID and MAGIC-generated perfusion maps. Each trial consisted of 20 sets of four perfusion maps (CBV, CBF, MTT, and TTP), where each set was a randomly selected real or synthetic perfusion series of one patient among the 20 patients. The real and synthetic sets of perfusion maps of the same patient were placed in different trials.
Furthermore, the patients were randomly shuffled and renamed using pseudo-IDs for anonymization within each trial. The shuffle was done so that the real and synthetic sets of perfusion maps from the same patients did not necessarily have the same indices in both trials. Each patient’s NCCT and CTP images were provided in one montage view, where five columns represented the NCCT image, CBV, CBF, MTT, and TTP maps, and the rows were the slices from different axial axis locations Fig. \ref{fig:double_blind_example}.
Seven experienced clinicians independently reviewed each image set from two trials conducted on different days to minimize confirmation bias. Four board-certified neuroradiologists (D.R., J.R., P.S., I.T) with  20, 31, 27, 22 years of experience, respectively, and three vascular neurologists (C.W., A.S., P.T.) with 16, 8, and 8 years of experience, respectively, evaluated the images. While neuroradiologists provide final formal interpretations in clinical workflows, vascular neurologists routinely interpret CTP studies in acute stroke settings and are often the first to make time-sensitive treatment decisions based on perfusion images. Recognizing their critical role in early stroke care and their extensive experience with perfusion imaging, we also included vascular neurologists in this validation process. The clinicians were asked to answer a series of questions on a questionnaire (Supplementary Fig. S.3). 

The questionnaire was designed in consultation with the three radiologists on the team (D.R., J.R., P.S) to assess four objectives on the diagnostic quality of perfusion maps generated by the MAGIC: (1) are the maps physiologically consistent; (2) are the maps diagnostically acceptable; (3) are the diagnosis consistent between the synthetic CTP images and the real CTP images; and (4) how confident are the raters in the given diagnosis to assess diagnostic confidence. Further statistical analysis was conducted on the physicians’ responses to the provided questionnaire, as presented in the results section.

\subsubsection{Significance Test for Overall Comparison}
As this work represents the first-of-its-kind deep learning approach to generate non-contrast CT perfusion maps from NCCT alone, the clinical evaluation focused on demonstrating that the perfusion maps generated using MAGIC from NCCT scans were diagnostically comparable to those produced by RAPID from contrast-enhanced CTP imaging data. To rigorously assess the consistency between the two methods, we evaluated inter-method agreement across four key diagnostic questions using both agreement percentage and Cohen’s kappa coefficient $\kappa$. While Cohen’s kappa is a widely used metric that accounts for the level of agreement expected by chance
Cohen’s kappa is a widely used metric that accounts for the level of agreement expected by chance, reporting both percent agreement and Cohen’s $\kappa$ is recommended to provide a more comprehensive assessment, particularly when response distributions are skewed or imbalanced \cite{ sim2005kappa, mchugh2012interrater, flight2015disagreeable}. The kappa coefficients and agreement percentages together with their 95\% confidence intervals were calculated to measure the agreement between MAGIC maps from NCCT scans and RAPID maps from contrast-enhanced CTP imaging data. We also conducted a power analysis to assess whether the sample size used in this study was sufficient to yield precise agreement estimates. This analysis using the empirical agreement metrics suggested that we would have greater than 80\% power to detect a kappa of 0.35 at the 0.05 significance level.

\section{Results}
\subsection{Performance in image quality and structural fidelity}
\begin{figure}[!htbp]
\centering
\includegraphics[trim=1cm 9cm 1cm 9cm, clip,width=\linewidth]{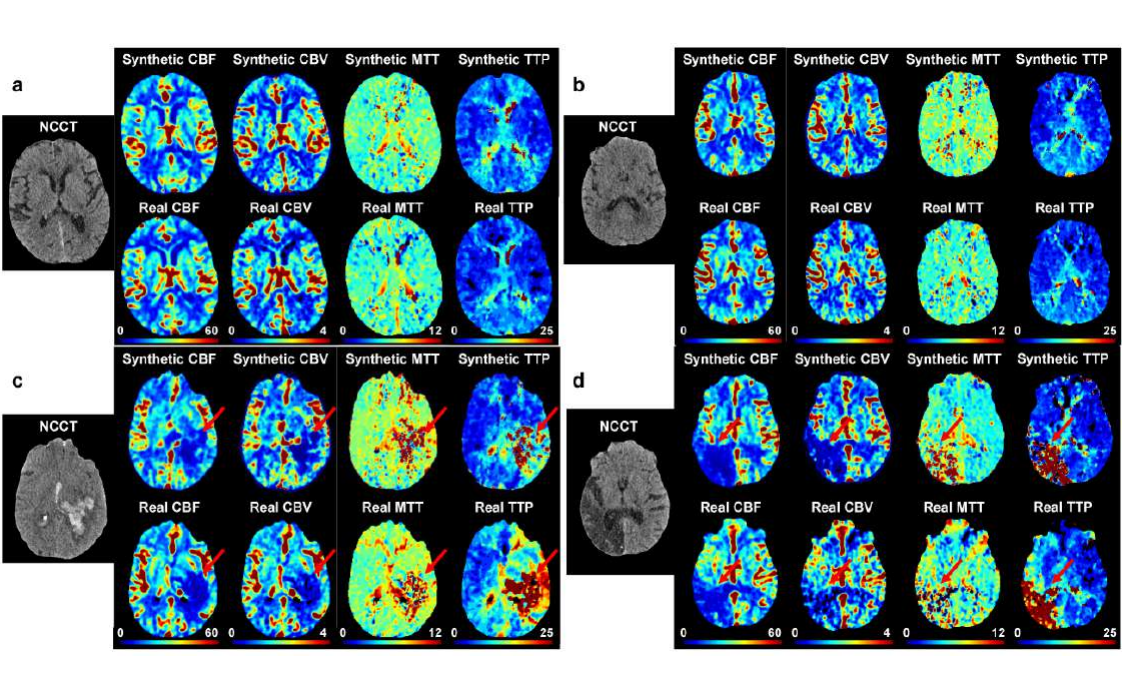}
\caption{A visual comparison of real CTP maps and synthetically generated maps via MAGIC shows efficacy in characterizing the brain’s hemodynamic activity from NCCT imaging alone. \textbf{a}. 75Y female patient presenting acute subdural hematoma layering along the right falx. Perfusion imaging shows increased TTP and increased CBV and CBF along the area of the acute subdural hemorrhage. Synthetic CTP is largely consistent with clinical CTP. \textbf{b}. 63Y male patient presenting patchy small vessel ischemic disease of the white matter. Perfusion imaging shows no focal perfusion defect. Synthetic CTP is consistent with clinical CTP, showing normal perfusion activity. \textbf{c}. 63Y male patient presenting left cerebral intraparenchymal hemorrhage. Perfusion imaging shows decreased CBF and CBV at the region of intraparenchymal hemorrhage with increased TTP throughout the left parietal, left posterior frontal, and left temporal lobes. Synthetic CTP is largely consistent with clinical CTP, showing elevated MTT and TTP with decreased CBF and CBV. \textbf{d}. 65Y male patient presenting encephalomalacia in the right parietal, occipital, and temporal lobes, compatible with prior infarcts. Perfusion imaging shows increased TTP with decreased CBF and CBV in the region of encephalomalacia. Synthetic CTP is consistent with trends in clinical CTP, showing elevated TTP and decreased CBV and CBF.}
\label{fig:comparison}
\end{figure}

Extensive quantitative and qualitative evaluations were performed on 168 patients who were completely independent of the training dataset. Among the 168 patients for evaluation, 104 have no infarction (62\%), 29 has mild to moderate infarction (17\%), and 35 has severe infarction (21\%). This distribution reflects the real distribution of infarction severity in the real-world dataset. To better reflect our model’s performance on mild to severe infarction patients, we oversampled the patients with infarction compared to the overall dataset which only has 7\% patients with mild to moderate infarction and 12\% patients with severe infarction. We applied the same preprocessing techniques, including skull stripping, slice extraction, and registration, to each patient’s NCCT and CTP series. Next, we use our MAGIC model to synthesize perfusion maps; MTT, TTP, CBF, and CBV, from each patient’s NCCT imaging. Fig. \ref{fig:comparison} depicts an example of a synthesized CTP image and its corresponding real CTP image. The synthesized perfusion imaging accurately characterizes both the style of each perfusion map type and the content encoded in that image.

\begin{figure}[!htbp]
\centering
\includegraphics[trim=0cm 7cm 0cm 6cm, clip,width=0.85\linewidth]{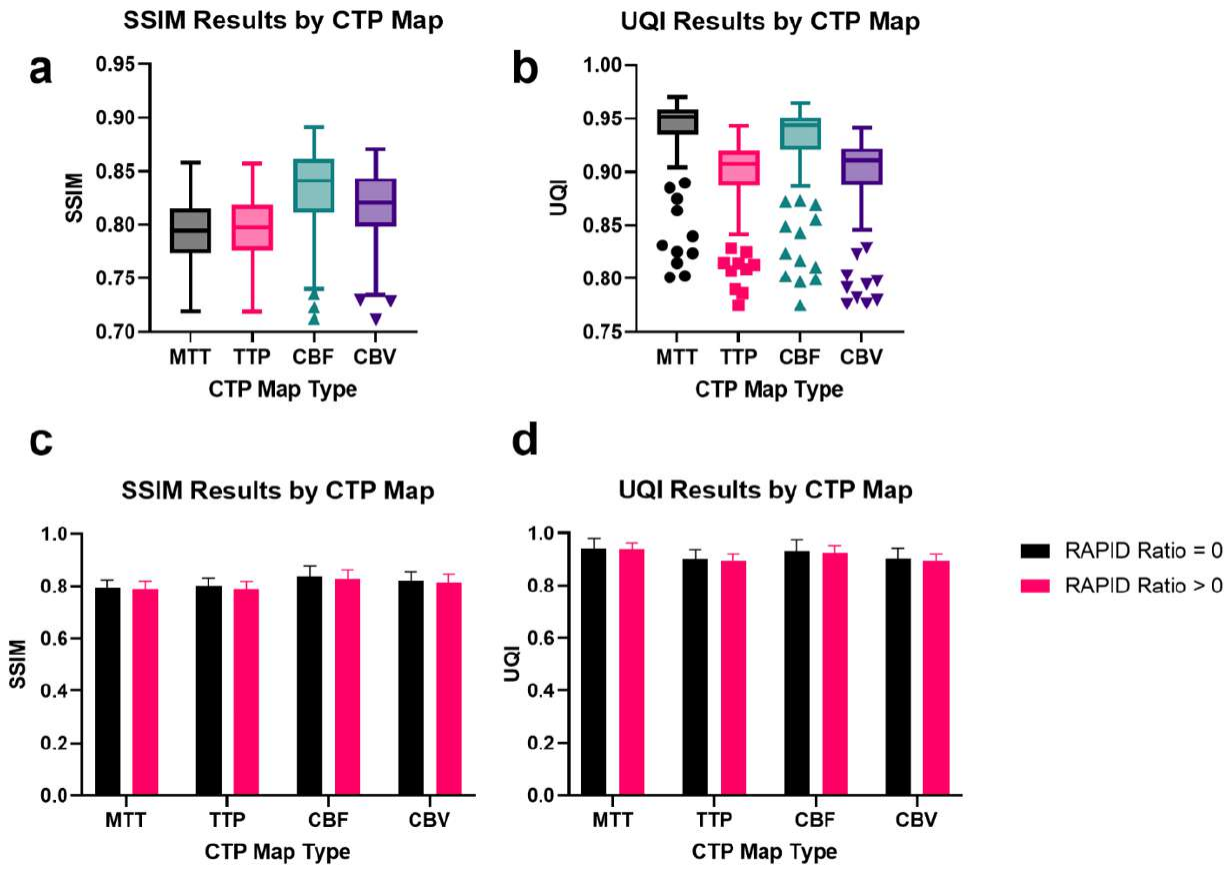}
\caption{MAGIC evaluated by SSIM and UQI separated by CTP map type and RAPID ratio. \textbf{a,b.} MAGIC results evaluated using SSIM and UQI metrics. MAGIC performed well in synthesizing structurally similar perfusion imaging across all map types. \textbf{c, d.} SSIM and UQI results stratified by the patient’s RAPID-calculated ratio of infarct core volume to tissue-at-risk volume. Among 168 test patients, 104 patients have a ratio = 0, and 64 patients have a ratio \(>\) 0. \textbf{c.} Mean SSIM values exceed 0.790 and \textbf{d.} mean UQI values exceed 0.890 across all map types, indicating very high structural integrity in the synthesized CTP imaging.}
\label{fig:quantitative_metrics}
\end{figure}

To assess the structural fidelity and similarity of the synthesized perfusion images to their corresponding real images, we used the Structural Similarity Index Metric (SSIM) and the Universal Quality Index (UQI). Clinically, SSIM and UQI are considered important metrics since they indicate structural fidelity and suggest similarity in the diagnostic quality of a target and a real image. The results, presented in box plots in Fig. \ref{fig:quantitative_metrics} A-B, show that our model achieved high fidelity synthesis of the real perfusion maps for all four types. Furthermore, we evaluated the SSIM and UQI values across all perfusion map types and infarction presentations and observed consistently high values (Fig \ref{fig:quantitative_metrics} A-B), indicating that our model accurately synthesized the structure of the real perfusion images.

Rather than striving for pixel-perfect replication of real CTP data, our synthesized imaging captures both global and local trends in the imaging that can offer crucial diagnostic information about a patient's brain perfusion activity. This approach is akin to commercial CTP software, which may not have pixel-to-pixel identity but still provide valuable diagnostic information. Our overall evaluation of the synthetic perfusion results produced by MAGIC indicates high levels of structural agreement with clinical perfusion imaging.

\subsection{Robustness to abnormal perfusion activity}
One crucial factor in assessing the performance of MAGIC is its ability to handle abnormal perfusion conditions, such as those observed in mild to severe ischemia. To evaluate its robustness under such pathological circumstances, we stratified our test set into two categories based on the core-to-penumbra ratio in RAPID (also known as RAPID ratio): RAPID ratio \(>\) 0 or RAPID ratio = 0. The results of this evaluation, presented in Fig. \ref{fig:quantitative_metrics}C-D and, demonstrate little deviation in the SSIM and UQI values across infarction classes. Our analysis indicates that MAGIC performs comparably well across different disease presentations, suggesting that it is highly resilient to abnormal perfusion activity.

\subsection{Physician-In-the-Loop (PILO) module integrates anatomical and hemodynamic information}
The PILO module was developed by combining the initial NCCT input with the penultimate CTP output in the final layers of the generator. This combined data is then passed through a final transposed convolution layer, which generates the synthetic perfusion output. In this module, physicians have the ability to adjust the weights of the transposed convolution layer, which can scale the level of hemodynamic and anatomical information present in the synthetic perfusion imaging.

The PILO module enables physicians to perform task-specific investigations without modifying the trained generator or influencing any quantitative results. The weights of the final deconvolutional layer of the PILO module can be scaled to either increase or decrease the perfusion information ratio. By changing the weights of the final deconvolutional layer, the perfusion information ratio can be adjusted, representing the balance between hemodynamic and anatomical information in the synthetic output.  Increasing the ratio increases the amount of hemodynamic information, while decreasing the ratio increases the amount of anatomical information. 

\begin{figure}[!htbp]
\centering
\includegraphics[trim=0cm 9cm 0cm 8cm, clip,width=0.85\linewidth]{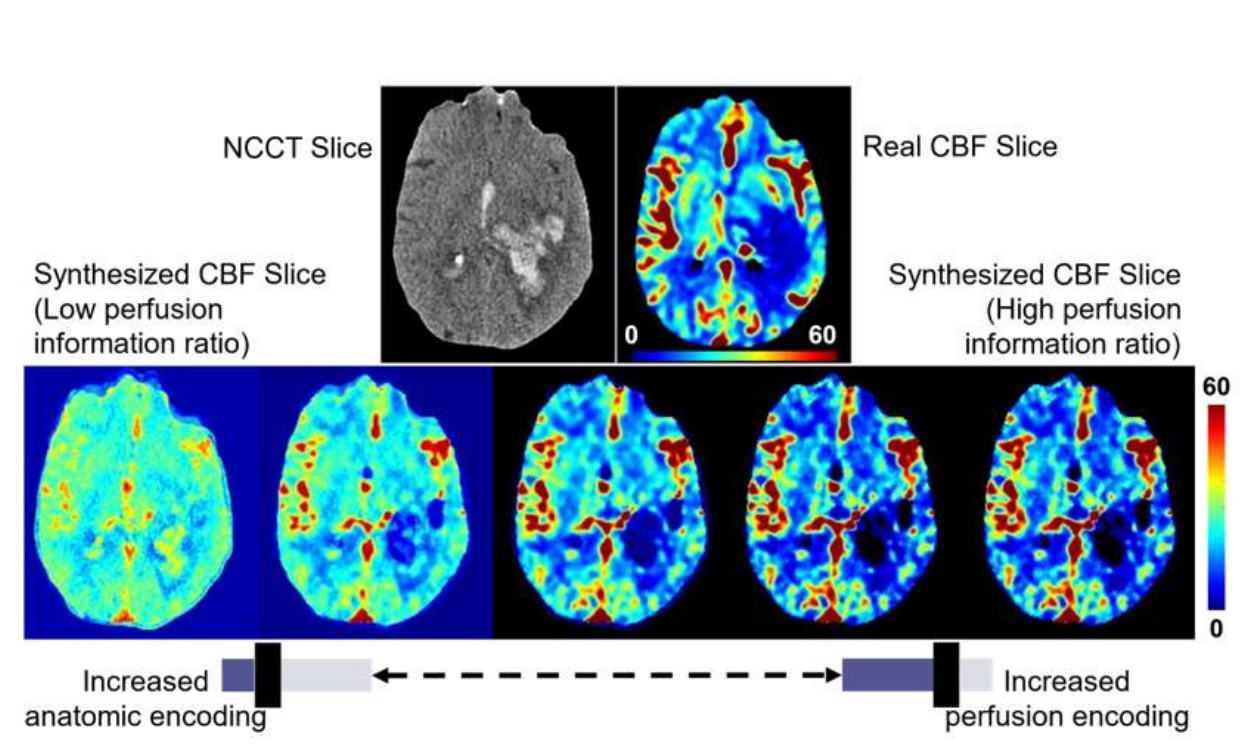}
\caption{Illustration of the effect of adjusting the perfusion information ratio in the PILO module to emphasize encoded perfusion activity versus anatomic representation in the synthesized maps. The presented image is from a 63Y male patient presenting left cerebral intraparenchymal hemorrhage. Perfusion imaging shows decreased CBF at the region of intraparenchymal hemorrhage. Increasing perfusion encoding shows exaggeration of the decrease in CBF due to the intraparenchymal hemorrhage. By adjusting the perfusion information ratio, the physician may choose to integrate more anatomical or perfusion information in the regions of interest from the synthesized perfusion maps for radiologist-preferred, task-specific, and patient-individualized analysis of the patient’s imaging.}
\label{fig:PILO}
\end{figure}

Physicians can leverage the PILO module as an exploratory tool to make informed clinical decisions regarding a patient's condition. The module allows for visualizing trends in the brain's perfusion activity by emphasizing the encoding of perfusion information. Fig. \ref{fig:PILO} demonstrates the effect of rescaling the ratio of weights in the PILO module for encoding hemodynamic and anatomic data. These results show the importance of the PILO module as an exploratory diagnostic tool for physicians to further investigate a patient's brain perfusion activity.

\subsection{MAGIC generates higher-fidelity and higher spatial resolution\\three-dimensional perfusion maps}
\begin{figure}[!htbp]
\centering
\includegraphics[trim=0cm 6cm 0cm 2cm, clip,width=\linewidth]{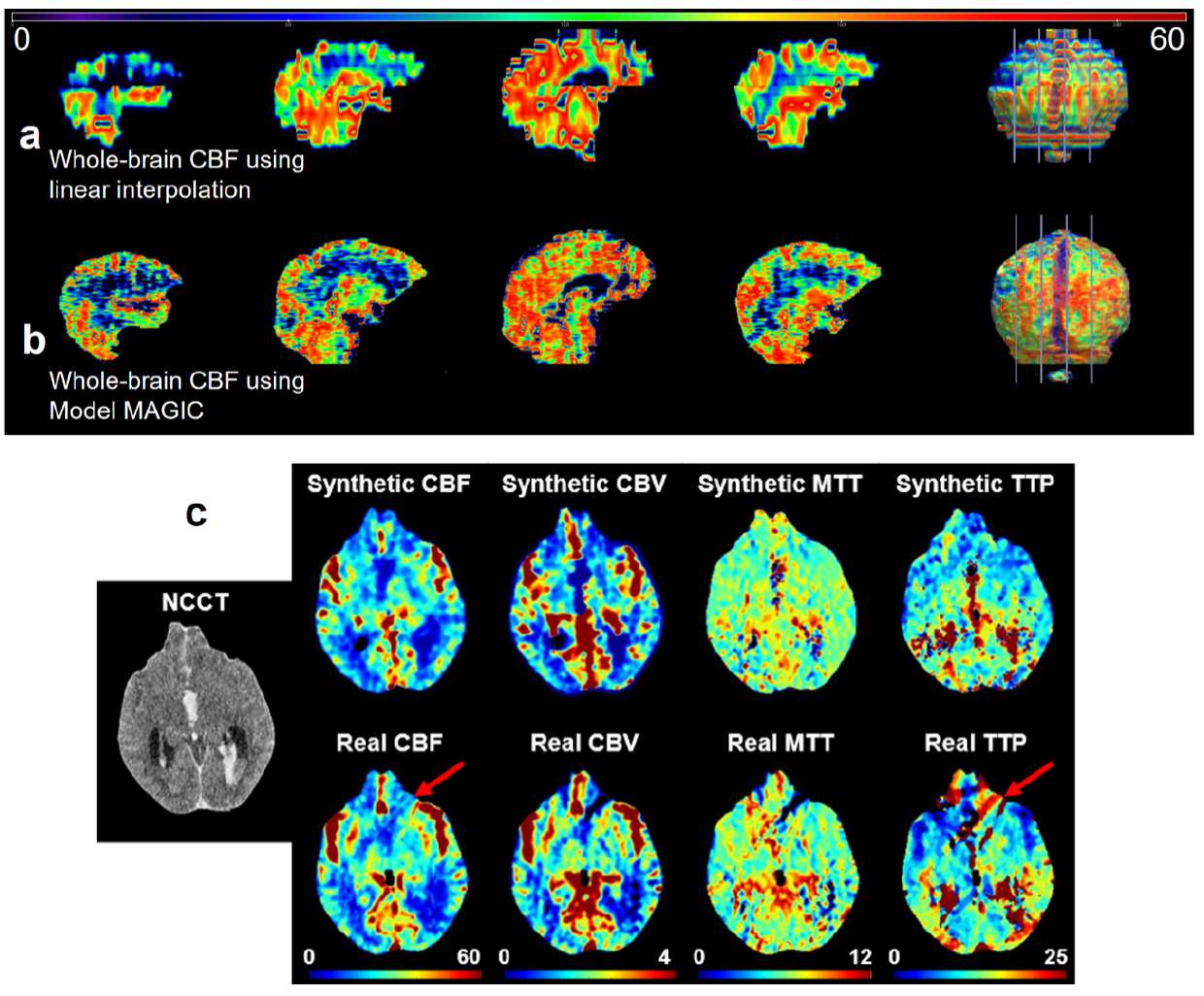}
\caption{\textbf{a.} Illustration of increased linear spatial density of whole-brain perfusion maps enabled by MAGIC. The whole-brain perfusion map is reconstructed using linear interpolation between 8 mm CBF map slices generated using RAPID CT perfusion analysis software. This results in a jagged and overly smoothed appearance that doesn’t maintain much detail of the brain’s structure. \textbf{b.} The whole-brain CBF map is reconstructed by applying MAGIC to an NCCT volume. The synthetic CBF volume maintains a higher level of detail and structural soundness at a higher spatial density. Important edge details and activity are preserved in visualizing whole-brain perfusion activity because we synthesize perfusion maps at each slice location. \textbf{c.} MAGIC can synthesize perfusion maps that are less susceptible to the presence of artifacts. For the given sample, the patient’s real perfusion imaging generated using RAPID contains streaking due to motion artifacts. However, the synthesized CTP maps via MAGIC do not contain these artifacts.}
\label{fig:3D}
\end{figure}

MAGIC has the capability to generate three-dimensional perfusion maps with high fidelity and spatial resolution for each slice of the NCCT input. The high spatial resolution axial-view perfusion maps have a resolution of 1mm,
compared with approximately 8 mm in commercial software\cite{kudo_differences_2010}.  This improved resolution allows for more informed clinical decisions, as demonstrated in Fig. \ref{fig:3D}A-B.

Additionally, MAGIC reduces the likelihood of motion artifacts in the generated perfusion maps. Compared to traditional perfusion maps, which require the administration of intravenous contrast agents and repeated imaging, MAGIC synthesizes perfusion imaging from NCCT imaging, which has a much shorter exposure time and is less susceptible to motion artifacts \cite{fahmi_head_2013}. This is illustrated in Fig. \ref{fig:3D}C. The ability to generate high-resolution, motion-artifact-free perfusion maps without the need for contrast agents presents a meaningful advantage of MAGIC over traditional CTP imaging.

\subsection{Agreement between MAGIC and RAPID}

Agreement between MAGIC and RAPID was assessed using contingency tables and visualized in agreement plots (Figure S.5) \cite{bangdiwala2013agreement}. The plots highlight the distribution of responses across categories for each question, showing where ratings agree or diverge. For instance, strong agreement is evident along the diagonal for most questions, while off-diagonal counts indicate mismatches. Furtheremore, Table 1 summarizes the agreement rates and corresponding kappa values across all questions. MAGIC and RAPID demonstrated high raw agreement rates on most questions—for example, 95\% agreement on Question 2B and over 90\% on Question 2A. However, some of the corresponding kappa values are low or even zero. In the case of Question 2B, although the agreement rate is very high (95\%), the $\kappa$ value is 0.00. This occurs because nearly all responses fall into the same category, leading to high expected agreement by chance. This highlights a known limitation of Cohen’s kappa, where the metric can under-estimate agreement in situations with imbalanced class distributions or low variability in ratings\cite{ sim2005kappa, mchugh2012interrater, flight2015disagreeable}. Despite this, several questions show meaningful agreement beyond chance, with the highest kappa values observed for Question 2C ($\kappa$ = 0.362) and Question 3 ($\kappa$=0.331). These results suggest that MAGIC enables clinicians to make diagnostic interpretations that are generally in line with those made using RAPID, further supporting its potential utility in clinical workflows without requiring contrast-based imaging. These findings demonstrate that despite some limitations in the kappa statistic due to skewed response distributions, MAGIC shows strong overall agreement with RAPID, supporting its clinical interpretability and robustness.

\begin{table}[ht]
\centering
\caption{Inter-rater agreement and 95\% confidence interval (CI) between MAGIC and RAPID from seven clinical evaluators: Cohens kappa that adjusts for chance agreement and percent agreement that reflects the proportion of exact matches in responses.}
\label{tab:kappa}
\begin{tabular}{lcc}
\hline
\textbf{Question} & \textbf{Agreement [95\% CI]} & \textbf{Cohens Kappa* [95\% CI]} \\
\hline
Question 1 & 80.71\% [73.39\%, 86.39\%] & 0.285 [0.081, 0.489] \\
Question 2A & 90.71\% [84.76\%, 94.49\%] & -0.019 [-0.042, 0.004] \\
Question 2B & 95.00\% [90.04\%, 97.56\%] & 0.000 [0.000, 0.000] \\
Question 2C & 75.71\% [67.99\%, 82.07\%] & 0.362 [0.207, 0.517] \\
Question 2D & 82.86\% [75.76\%, 88.20\%] & 0.035 [-0.119, 0.1885] \\
Question 3 & 53.24\% [44.97\%, 61.33\%] & 0.334 [0.223, 0.445] \\
Question 4 & 55.00\% [46.74\%, 63.00\%] & 0.298 [0.176, 0.421] \\
\hline
\end{tabular}
\small
 *Lower or zero Kappa values indicates cases where Cohen's kappa is affected by marginal imbalance \cite{ sim2005kappa, mchugh2012interrater, flight2015disagreeable}.
\end{table}

\subsection{Clinical evaluation and diagnostic validation}
To assess the clinical efficacy of MAGIC, a double-blinded study was conducted with seven clinicians independently evaluating 20 out of the 168 patient scans. Among these 20 patients, seven showed regular cerebral perfusion activity without any infarction, six showed mild to moderate infarction, and seven showed severe infarction. The clinicians evaluated the scans based on four criteria, namely authenticity, diagnostic quality, diagnosis, and diagnostic confidence, all of which were directly linked to clinical diagnosis outcomes. The study was split into two groups, with the clinicians receiving either synthetic CTP or real CTP for each patient, and a random shuffle was conducted to determine which CTP type was included in each trial. To ensure blinding, all perfusion maps were shuffled before presentation, so neither the researchers nor the clinicians knew whether any given image was real (RAPID) or synthetic (MAGIC). A representative sample of the blinded images shown to clinicians is provided in Fig. \ref{fig:double_blind_example}.

The clinical evaluation revealed that the CTP maps produced by MAGIC and RAPID had statistically comparable performances in terms of clinical authenticity, as shown in Table. S.1. Despite different clinician-specific preferences, all raters were consistent in their classification rates across both groups of  RAPID and MAGIC perfusion maps. Additionally, individual and aggregated responses to Question 2—“What is the diagnostic quality of this perfusion map?”—are presented across Tables S.2 to S.5. These tables show the number of responses categorized as Acceptable, Indeterminate, or Unacceptable for four perfusion map types (CBV, CBF, MTT, and TTP) generated by either the MAGIC or RAPID pipeline. The overall distribution pooled across all seven doctors, as well as individual doctor distributions, indicate that most images were rated Acceptable across both methods and all map types. Indeterminate and Unacceptable ratings were infrequent. This pattern was generally consistent across raters, though some variability was observed, particularly for MTT maps where certain doctors (e.g., Doctor 2 and Doctor 4) reported more Indeterminate or Unacceptable ratings for MAGIC images compared to RAPID. Notably, Doctor 6 rated all maps from both methods as Acceptable, showing no variability in their assessments. Overall, these results support the conclusion that MAGIC-generated synthetic perfusion maps are comparable in diagnostic quality to those produced by the clinical standard RAPID pipeline. To further assess the diagnostic utility of synthetic images, we analyzed the agreement in clinical ratings between RAPID and MAGIC-generated CTP maps. As shown in Tables S.6 and S.7, diagnostic responses for Question 3 (diagnosis; categories 1–4) and Question 4 (diagnostic confidence; categories 1–5) were largely concentrated along the diagonal, indicating strong agreement in both diagnostic outcomes and confidence levels between real and synthetic images. For example, 43 cases were rated as category 1 by both methods, and confidence ratings showed similar consistency, with 44 cases rated as 4 and 25 cases rated as 5 by both methods.

In addition to the numerical evaluation, three experienced radiologists also provided written descriptions of their evaluation of the differences between the real and synthetic perfusion maps. According to their observations, the non-contrast CT (NCCT) image was the fundamental imaging used to start working on the CTP images. They found that the CBF and CBV images for the real and synthetic images were close enough to have the same diagnostic outcome. Occasionally, the MTT and TTP images were very different, but with correlative imaging of non-contrast CT, CBV, and CBF, the clinical diagnosis was achieved. Therefore, they found no meaningful difference in the diagnostic ability in both image sets. When looking for differences between the synthetic and post-contrast (real) CTP studies, one radiologist said they look for differences in degree of pixelation and in anatomic conformation to features seen on the non-contrast CT and the other parameters. However, in general, they found that their ability to differentiate the synthetic from the non-synthetic was not great.

In summary, MAGIC shows comparable diagnostic performance to commercial software and has the potential to improve the clinical door-to-needle time for acute stroke patients. The use of MAGIC can lead to a faster, cheaper, and safer triage and diagnosis process for acute stroke, which can ultimately save lives and reduce healthcare costs.

\subsection{Ablation studies and Comparative Analysis}

This section presents ablation studies to evaluate the performance of MAGIC, followed by a comparison with state-of-the-art models, including Pix2Pix and U-Net.  The MAGIC model consistently outperforms nearly all other models, achieving the highest SSIM and UQI values across all metrics (Table S.8). However, when proposed multimodal and extrema loss are removed from the MAGIC model, its performance declines. For instance, in CBF, the SSIM and UQI for MAGIC are 0.8302 and 0.9313, respectively, but removing the multimodal loss reduces these values to 0.7911 and 0.6129. Similarly, removing the extrema and both losses also leads to a noticeable degradation in performance. Similar trends are observed across other perfusion parameters such as CBV, MTT, and TTP. This highlights that both losses play a critical role in enhancing the structural fidelity and perceptual quality of the generated perfusion maps. Furthermore, we compared MAGIC against two with state-of-the-art deep learning models: U-Net and Pix2Pix (Table S.9). While U-Net generally outperforms Pix2Pix, particularly in terms of UQI (e.g., 0.9093 for CBV and 0.8600 for TTP), MAGIC consistently delivers the best overall quality. For instance, MAGIC outperforms both U-Net and Pix2Pix in SSIM and UQI on the CBF map, with a notable UQI of 0.9313. Even for maps where U-Net performs well (e.g., MTT and TTP), MAGIC maintains superior quality, highlighting its robustness across modalities. Overall, MAGIC generates higher-quality perfusion outputs, making it a more reliable tool for downstream clinical interpretation and analysis.


\section{Discussion}
\begin{figure}[!htbp]
\centering
\includegraphics[trim=14cm 10cm 14cm 8cm, clip,width=0.99\linewidth]{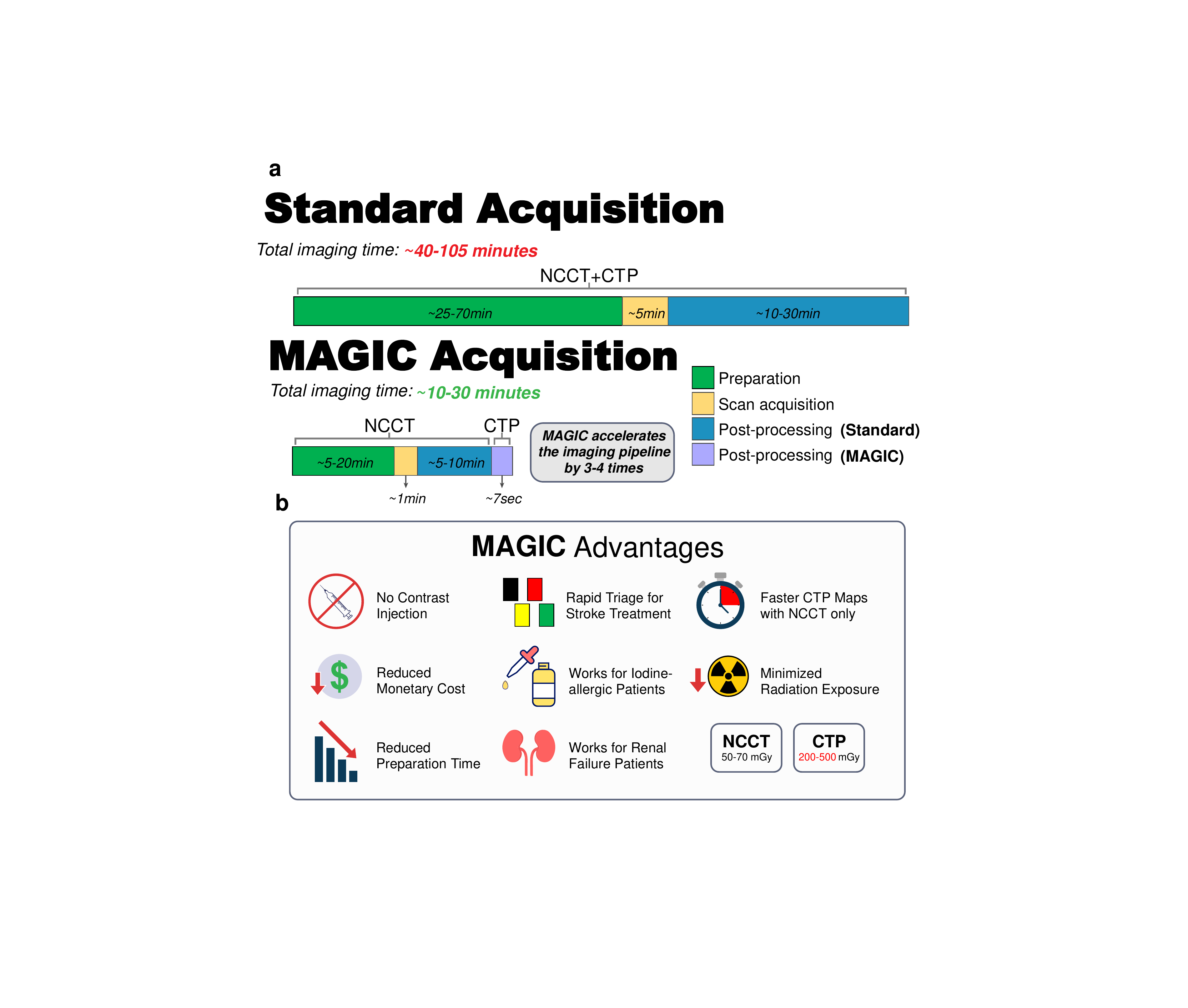}
\caption{Proposed advantages of MAGIC and comparison against existing workflows. \textbf{a.} The amount of time from arriving to acquiring CT perfusion maps is considerably reduced. Preparation and scan acquisition for CT perfusion is cut out from the MAGIC pipeline. Times are estimated from existing workflows. \textbf{b.} MAGIC removes the preparation and resources necessary to acquire CT perfusion maps. This pipeline reduces the number of CT perfusion contraindications. Removal of contrast injection allows for rapid triage for stroke treatment, faster CT perfusion acquisition with only NCCT, reduced monetary cost, minimized radiation exposure, and reduced preparation time.}
\label{fig:advantages}
\end{figure}

The proposed MAGIC model, including the novel physiology-informed loss terms and physicians-in-the-loop module, performs favorably in comparison to commercial software for perfusion imaging with contrast. MAGIC enables rapid, contrast-free, high fidelity, high resolution, and physician-preferred generation of CTP imaging from only the NCCT series. Implementing this model in clinical practice has the potential to: reduce the need for contrast injection, produce diagnostically comparable perfusion imaging, reducing acquisition time (roughly 90 seconds or more using contrast-enhanced CTP \cite{kasasbeh_optimal_2016} to roughly 1.5 seconds using NCCT only \cite{singh_time_2018}), forgoing the length of preparation time for contrast injection, improving scan-to-map post-processing speed (roughly 5 minutes using RAPID software 31 to about 7.43 sec per volume of 161 slices once trained), providing cost-effectiveness at which physicians can assess a patient’s hemodynamic activity, and saving critical time at rapid triage stage (Fig. \ref{fig:advantages})\cite{rapid_nodate}. This can further be used to enable CTP imaging in limited-resource hospitals that currently lack the proper training, staff, contrast agent, funds, or facilities to conduct CTP protocols. Additionally, the use of non-contrast imaging in synthesizing perfusion data via MAGIC presents potential benefits in reducing the radiation dose of routine imaging procedures. The radiation output of a head CTP exam is roughly 4 to 10 times that of a head NCCT exam (200 to 500 mGy in CTP vs. 50 to 70 mGy in NCCT). Prior studies have raised the concern of patient harm from radiation exposure during CTP, especially within the context of multimodal CT protocol\cite{zensen_radiation_2021,mnyusiwalla_radiation_2009}. The use of non-contrast imaging in the synthesis of CTP imaging also circumvents the potential negative side effects associated with iodinated contrast injection including: allergic reactions, cardiac and renal failure, and pulmonary edema \cite{bottinor_adverse_2013}. Additionally, the elimination of intravenous contrast injection can substantially reduce the supply cost. These improvements to the safety, efficiency, cost, and accessibility of CTP imaging positions MAGIC as a valuable asset in enabling contrast-free perfusion imaging. 

This paper is the first to address the research question and clinical potential of a contrast-free generation of CTP imaging, using image-to-image translation techniques between the NCCT and CTP series. We introduced cutting edge innovations into this physiology-informed multitask deep learning model. The novel physicians-in-the-loop (PILO) module provides radiologists with an interactive and adaptive diagnostic tool to select and visualize the hemodynamic activity embedded in anatomical imaging. The radiologist evaluators in this work stated that: PILO is a valuable tool that allowed for better visualization and diagnosis of the brain’s perfusion activity than the clinical perfusion scans alone.

Our evaluation demonstrates that MAGIC can utilize deep learning-based multimodal image translation techniques to synthesize non-contrast perfusion maps. Our synthetic perfusion imaging has comparable structural fidelity and diagnostic quality to real perfusion imaging generated using commercial software. Perfusion maps generated by MAGIC, especially CBF and CBV maps, displayed accurate hemodynamic status present in real imaging. There was no meaningful difference in the values of the image metrics across all four image map classes, which indicates that MAGIC performs stably across perfusion map types. Additionally, there was no meaningful difference between the performance metrics across different patients with varying levels of infarction status, which indicates that MAGIC is robust to abnormal hemodynamic activities such as acute ischemia.

Our double-blinded study revealed that our synthetic perfusion imaging accurately portrays real imaging. Additionally, our synthetic perfusion imaging has no distinguishable features that the clinicians could use to consistently discriminate between synthesized and real imaging. The clinicians evaluators in this study concluded that the synthetic imaging was of sufficient diagnostic quality to confidently diagnose a patient and generate a treatment plan. In concurrent discussions with the clinicians, we found that they prefer to use CBV and TTP imaging to produce a diagnostic decision and characterize the volumes of the infarcted core and the ischemic penumbra. We found that the diagnoses and treatment plans the clinicians produced using the synthesized imaging were statistically consistent with those they gave to the corresponding real perfusion imaging. Furthermore, although a mixed-effects analysis could further quantify reader-level variability, the current sample of ratings per doctor was not sufficient to support a stable generalized linear mixed model; therefore, we relied on descriptive agreement patterns for this study. 

The diagnostic accuracy and computational efficiency offered by MAGIC potentiates more rapid and cost-efficient clinical workflows in assessing neurovascular health and treating acute ischemia. Meanwhile, there are some limitations of this study. Since non-contrast CTP synthesis via deep learning image translation is a novel research problem, there are no baseline approaches to which we can compare our results. Additionally, we found that the MTT imaging produced by the RAPID software was consistently ranked as “unacceptable” when evaluated by clinicians. The RAPID data have limitations that account for its current limited use in the clinical setting. MAGIC is unable to capture a complete representation of microvascular activity consistently, even though it performs well in predicting hemodynamic activity from non-contrast imaging. For example, the current MAGIC iteration cannot predict small-vessel occlusions based on non-contrast imaging alone (Supplementary Fig. 7). This limitation is evident in cases where low-density areas in the periventricular white matter, typically indicative of small-vessel ischemic injury, are not captured by the perfusion parameters. We plan to address these issues in future improvements of MAGIC by incorporating more detailed perfusion imaging data from other cohorts into the training set. 

Furthermore, we aim to address the issues present with capturing the activity from the temporal perfusion maps, TTP and MTT, through further iterations on our methodology and diversity of our training set. A more accurate encoding of the temporal map classes will enable clinicians to make a more informed diagnosis. While the MAGIC model has shown promising results, its generalizability to external datasets remains a key area for future work. It has not yet been tested on external data, which may vary in clinical settings, imaging protocols, and patient demographics. Evaluating the model on diverse, independent datasets is essential for assessing its robustness and clinical applicability. Additionally, to improve synthetic CTP map generation, we plan to incorporate advanced models like diffusion models, which could enhance the accuracy and consistency of temporal maps and improve microvascular activity predictions, ultimately increasing the clinical utility of the generated maps \cite{ rombach2022high, bae2024conditional, kazerouni2023diffusion, khader2023denoising, sun2025conditional}.
Furthermore, NCCT and CTP slices in the training or test sets may not be registered perfectly due to patient movement between scanning protocols while collecting imaging data. Our training framework requires paired and co-registered data between the NCCT and CTP imaging series. Considering this, we must only use the NCCT image slices with a corresponding CTP slice at the exact vertical location, since the axial CTP image slices are sparser in spatial density than the axial NCCT image slices. As a result, we could not incorporate our entire dataset into our training algorithm to fully leverage information from all slices in NCCT. MAGIC is still capable of learning the mapping between NCCT and CTP well and its ability to generalize to new data, even with the reduced training data size. Finally, although MAGIC can infer perfusion patterns from NCCT, it lacks interpretability to identify the NCCT features that drive these predictions. Therefore, MAGIC should be viewed as learning statistical associations rather than explicit physiological mechanisms.

Despite these limitations, our overall conclusion has been supportive that MAGIC exhibits impressive performance in capturing and characterizing the brain’s hemodynamic activity in various cases based solely on non-contrast, structural imaging. Our algorithm could be trained to generate perfusion maps in collaboration with specific vendors and perfusion software to meet the needs of diagnosis using perfusion imaging in particular organs and diseases. Additionally, we aim to implement MAGIC in a clinical setting to efficiently generate perfusion imaging for rapid assessment of patient neurovascular health. There is a trend of rapid stroke diagnosis based on non-contrast imaging alone \cite{wang_deep_2021}. Our MAGIC model will empower the development of a rapid computer-aided diagnostic system that can recommend treatment protocols for patients based on non-contrast CT imaging. 

In conclusion, the proposed MAGIC can generate contrast-free brain perfusion imaging from solely non-contrast, structural imaging, which exhibits comparable analytical and diagnostic results to the real perfusion imaging. As a result, this model’s implementation has the potential to expand the accessibility of CTP brain imaging while reducing the time, cost, and patient risk associated with these diagnostic protocols.


\section*{CRediT Authorship Contribution Statement}

\textbf{Wasif Khan}: Formal analysis, Investigation, Validation,  Writing - Original Draft, Writing - Review and Editing, Visualization, Software
\textbf{Kyle B. See}: Methodology, Formal analysis, Investigation, Writing - Original Draft, Writing - Review and Editing, Visualization, Supervision, Project Administration
\textbf{Simon Kato}: Methodology, Investigation, Data curation, Writing - Original Draft, Writing - Review and Editing, Visualization
\textbf{Ziqian Huang}: Software
\textbf{Amy Lazarte}: Data curation
\textbf{Kyle Douglas}: Data curation
\textbf {Xiangyang Lou} Formal analysis, Validation
\textbf {Teng J. Peng} Validation, Writing - Review and Editing
\textbf {Dhanashree Rajderkar} Validation
\textbf {John Rees} Validation
\textbf{Pina Sanelli} Validation
\textbf {Amita Singh} Validation
\textbf {Ibrahim Tuna} Validation
\textbf {Christina A. Wilson} Validation
\textbf{Ruogu Fang}: Conceptualization, Methodology, Formal analysis, Investigation, Resources, Data curation, Writing - Original Draft, Writing - Review and Editing, Supervision, Project administration, Funding acquisition. All authors contributed critical revisions to the manuscript and approved the final manuscript.


\section*{Competing Interests}
The authors declare that the research was conducted in the absence of any commercial or financial relationships that could be construed as a potential conflict of interest.


\section*{Acknowledgements}
This work was partially supported by the National Science Foundation (1908299) and its REU Supplement (2015326, 2128099). We acknowledge the University of Florida Integrated Data Repository (IDR) and the UF Health Office of the Chief Data Officer for providing the analytic data set for this project. Additionally, the Research reported in this publication was supported by the National Center for Advancing Translational Sciences of the National Institutes of Health under University of Florida Clinical and Translational Science Awards UL1TR000064 and UL1TR001427. The content of this publication, presentation and/or proposal is solely the responsibility of the authors and does not necessarily represent the official views of the National Institutes of Health. We thank Dr. Li Chen for his suggestion on statistical tests on sample size calculation. 


\section*{Data availability statement}
The data that support the findings of this study are available from the corresponding author [R.F.] upon reasonable request. The data are not publicly available due to patient data confidentiality.


\bibliographystyle{unsrt}
\bibliography{sample}



\section*{Supplementary material}

# Supplementary Materials

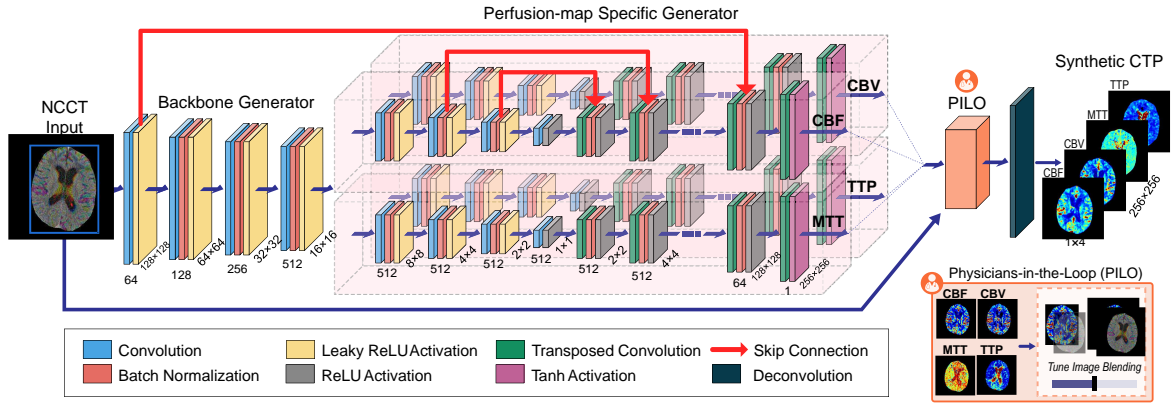

Figure S1: A diagram of the generator network architecture. The generator follows a modified U-Net architecture, with various network layers connected by skip connections. The first encoding layers share the encoded weights of each map before diverging into multi-task perfusion map generation. The physicians-in-the-loop module comprises the final layers of the architecture in which the NCCT input is concatenated with the initial perfusion output. A final deconvolutional layer is applied to generate the final perfusion map outputs.

Table S1: The results of Q1 from the double-blinded questionnaire: "Do you think this is a real CT perfusion map?". All raters were unable to differentiate between real and synthetic CTP imaging. Overall, both real (RAPID) and synthetic (MAGIC) CTP images were predominantly classified as "RAPID"— 117 and 118 responses, respectively. This trend held across individual doctors, with the exception of Doctor 7, who more frequently identified images as MAGIC. These findings suggest that raters predominantly treated both real and synthetic CTP maps as real, indicating that MAGIC is capable of generating synthetic CTP maps that closely resemble real ones.

| Doctor | RAPID | | MAGIC | |
| --- | --- | --- | --- | --- |
| | Is RAPID | Is MAGIC | Is RAPID | Is MAGIC |
| 1 | 12 | 8 | 15 | 5 |
| 2 | 19 | 1 | 19 | 1 |
| 3 | 19 | 1 | 16 | 4 |
| 4 | 20 | 0 | 18 | 2 |
| 5 | 20 | 0 | 18 | 2 |
| 6 | 19 | 1 | 19 | 1 |
| 7 | 8 | 12 | 13 | 7 |
| **Sum** | **117** | **23** | **118** | **22** |

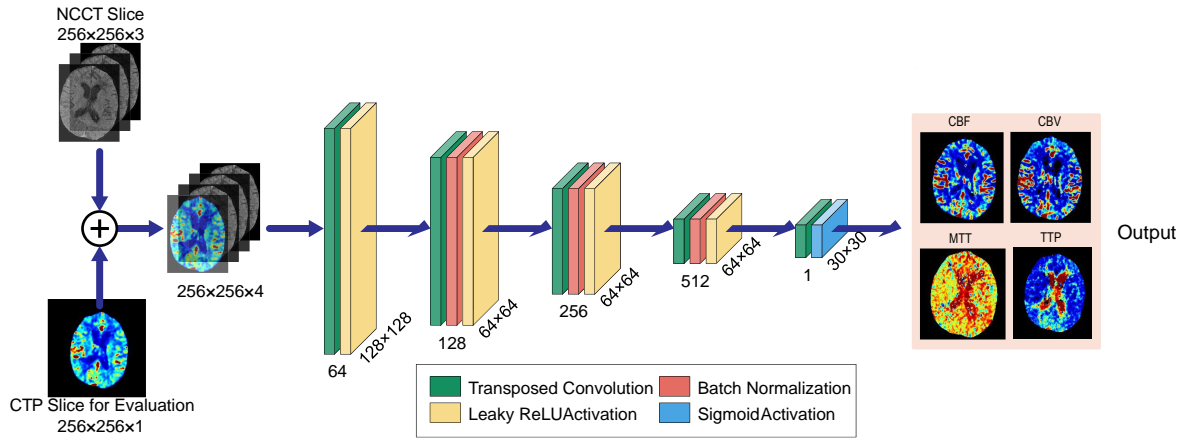

Figure S2: Discriminator network architecture which is based on the PatchGAN framework, which evaluates the realism of local image patches rather than entire images. The discriminator applies a series of transposed convolutional layers to the concatenated input of the NCCT slice and the CTP slice for evaluation.

Table S2: Comparison of individual responses to RAPID and MAGIC CTP for Questions 2: "What is the diagnostic quality of this perfusion map?" The table shows the response counts (Acceptable, Indeterminate, Unacceptable) for CBF maps processed by RAPID and MAGIC. Most images were rated "Acceptable" across both methods, with very few "Indeterminate" or "Unacceptable" responses. Acc = Acceptable, Indet = Indeterminate, Unacc = Unacceptable.

| Doctor | RAPID CBF | | | MAGIC CBF | | |
| --- | --- | --- | --- | --- | --- | --- |
| | Acc | Indet | Unacc | Acc | Indet | Unacc |
| 1 | 20 | 0 | 0 | 19 | 1 | 0 |
| 2 | 20 | 0 | 0 | 19 | 1 | 0 |
| 3 | 20 | 0 | 0 | 20 | 0 | 0 |
| 4 | 20 | 0 | 0 | 18 | 2 | 0 |
| 5 | 20 | 0 | 0 | 17 | 3 | 0 |
| 6 | 20 | 0 | 0 | 20 | 0 | 0 |
| 7 | 20 | 0 | 0 | 20 | 0 | 0 |
| **Sum** | **140** | **0** | **0** | **133** | **7** | **0** |

Table S3: Response comparison for RAPID and MAGIC CTP on Question 2: "What is the diagnostic quality of this perfusion map?" Counts of Acceptable (Acc), Indeterminate (Indet), and Unacceptable (Unacc) ratings for CBV maps are shown. Most images were rated "Acceptable" for both methods.

| Doctor | RAPID CBV | | | MAGIC CBV | | |
| --- | --- | --- | --- | --- | --- | --- |
| | Acc | Indet | Unacc | Acc | Indet | Unacc |
| 1 | 20 | 0 | 0 | 19 | 1 | 0 |
| 2 | 20 | 0 | 0 | 18 | 1 | 1 |
| 3 | 18 | 2 | 0 | 18 | 1 | 1 |
| 4 | 20 | 0 | 0 | 18 | 2 | 0 |
| 5 | 20 | 0 | 0 | 17 | 1 | 2 |
| 6 | 20 | 0 | 0 | 20 | 0 | 0 |
| 7 | 20 | 0 | 0 | 19 | 0 | 1 |
| **Sum** | **138** | **2** | **0** | **129** | **6** | **5** |

Q1. Do you think this is a real CT perfusion map?
   1. Yes
   0. No

Q2. For the following questions A-D, make your best efforts to choose "Acceptable" or "Unacceptable." If you choose "Indeterminate," please provide a short explanation of your answer.
   a. What is the diagnostic quality of the CBV maps?

      1. Acceptable
      0. Indeterminate
      -1. Unacceptable
   b. What is the diagnostic quality of the CBF maps?

      1. Acceptable
      0. Indeterminate
      -1. Unacceptable
   c. What is the diagnostic quality of the MTT maps?

      1. Acceptable
      0. Indeterminate
      -1. Unacceptable
   d. What is the diagnostic quality of the TTP maps?

      1. Acceptable
      0. Indeterminate
      -1. Unacceptable

Q3. What diagnosis would be given to this patient based on their NCCT and CTP imaging?
   1. Normal exam with no perfusion deficit
   2. Core infarct is less than or equal to 20% of the tissue at risk
   3. Core infarct is strictly more than 20% of the tissue at risk
   4. Core infarct is approximately equal to tissue at risk

Q4. Rate your confidence in the above diagnosis.
   1. Not confident
   2. Slightly confident
   3. Somewhat confident
   4. Fairly confident
   5. Completely confident

Figure S3: The questionnaire provided to evaluators for double-blinded evaluation.

Table S4: Comparison of responses for RAPID and MAGIC CTP on Question 2: Counts of Acceptable (Acc), Indeterminate (Indet), and Unacceptable (Unacc) ratings for MTT maps. Most images were rated "Acceptable" for both methods.

| Doctor | RAPID MTT | | | MAGIC MTT | | |
|---|---|---|---|---|---|---|
| | Acc | Indet | Unacc | Acc | Indet | Unacc |
| 1 | 20 | 0 | 0 | 20 | 0 | 0 |
| 2 | 14 | 0 | 6 | 7 | 4 | 9 |
| 3 | 18 | 2 | 0 | 19 | 1 | 0 |
| 4 | 4 | 2 | 14 | 0 | 2 | 18 |
| 5 | 20 | 0 | 0 | 18 | 2 | 0 |
| 6 | 20 | 0 | 0 | 20 | 0 | 0 |
| 7 | 17 | 0 | 3 | 17 | 0 | 3 |
| **Sum** | **113** | **4** | **23** | **101** | **9** | **30** |

Table S5: Comparison of responses for RAPID and MAGIC CTP on Question 2: Counts of Acceptable (Acc), Indeterminate (Indet), and Unacceptable (Unacc) ratings for TTP maps. Most images were rated "Acceptable" with few "Indeterminate" or "Unacceptable" responses.

| Doctor | RAPID TTP | | | MAGIC TTP | | |
|---|---|---|---|---|---|---|
| | Acc | Indet | Unacc | Acc | Indet | Unacc |
| 1 | 19 | 1 | 0 | 19 | 1 | 0 |
| 2 | 18 | 2 | 0 | 15 | 4 | 1 |
| 3 | 20 | 0 | 0 | 18 | 2 | 0 |
| 4 | 14 | 5 | 1 | 14 | 6 | 0 |
| 5 | 20 | 0 | 0 | 18 | 2 | 0 |
| 6 | 20 | 0 | 0 | 20 | 0 | 0 |
| 7 | 18 | 0 | 2 | 20 | 0 | 0 |
| **Sum** | **129** | **8** | **3** | **124** | **15** | **1** |

Table S6: Results for responses to Q3: "What diagnosis would be given to this patient based on their NCCT and CTP imaging?" The matrix on the left shows the distribution of diagnostic ratings (1–4) assigned to each case using real CTP (RAPID) versus synthetic CTP (MAGIC). The majority of responses are concentrated along the diagonal, indicating high agreement between diagnoses made with real and synthetic CTP images. For example, 43 cases were rated as category 1 by both methods, and similar agreement is seen for other categories.

| Method | | MAGIC | | | |
|---|---|---|---|---|---|
| | Option | 1 | 2 | 3 | 4 |
| | 1 | 43 | 14 | 0 | 2 |
| RAPID | 2 | 13 | 9 | 2 | 6 |
| | 3 | 3 | 2 | 6 | 13 |
| | 4 | 6 | 2 | 1 | 16 |

Table S7: Results for responses to Q4: "Rate your confidence in the above diagnosis." The matrix on the right displays the confidence ratings (1–5) for each case, comparing real and synthetic CTP. Most responses are clustered along the diagonal, especially at higher confidence levels (e.g., 44 cases rated as 4 by both methods, and 25 cases rated as 5 by both), indicating that raters' confidence in their diagnoses was generally consistent between RAPID and MAGIC CTP images.

| Method | | MAGIC | | | | |
|---|---|---|---|---|---|---|
| | Option | 1 | 2 | 3 | 4 | 5 |
| | 1 | 1 | 0 | 1 | 0 | 0 |
| | 2 | 0 | 3 | 1 | 4 | 2 |
| RAPID | 3 | 0 | 1 | 4 | 8 | 1 |
| | 4 | 0 | 9 | 6 | 44 | 15 |
| | 5 | 0 | 1 | 3 | 11 | 25 |

Table S8: Ablation study results for the MAGIC model across four perfusion maps: Cerebral Blood Flow (CBF), Cerebral Blood Volume (CBV), Mean Transit Time (MTT), and Time to Peak (TTP). SSIM and UQI scores are reported for the full MAGIC model and for variants with specific loss components removed (no multimodal loss, no extrema loss, and no loss components). The results highlight the importance of each loss in contributing to the final performance.

| Model | CBF | | CBV | | MTT | | TTP | |
|---|---|---|---|---|---|---|---|---|
| | SSIM | UQI | SSIM | UQI | SSIM | UQI | SSIM | UQI |
| No-multimodal | 0.7911 | 0.6129 | 0.7861 | 0.6222 | 0.8103 | 0.7422 | 0.79716 | 0.7203 |
| No-both | 0.7936 | 0.6186 | 0.7886 | 0.6243 | **0.8130** | 0.7442 | 0.80007 | 0.7234 |
| No extrema | 0.7835 | 0.6148 | 0.7807 | 0.5454 | 0.8036 | 0.7529 | 0.7997 | 0.7739 |
| **MAGIC** | **0.8302** | **0.9313** | **0.8158** | **0.9037** | 0.7938 | **0.9453** | 0.7975 | **0.9066** |

Table S9: Comparison of MAGIC with state-of-the-art models (U-Net and Pix2Pix) across CBF, CBV, MTT, and TTP.

| Model | CBF | | CBV | | MTT | | TTP | |
|---|---|---|---|---|---|---|---|---|
| | SSIM | UQI | SSIM | UQI | SSIM | UQI | SSIM | UQI |
| UNet | 0.6164 | 0.7747 | 0.7951 | **0.9093** | 0.8308 | 0.8783 | 0.8105 | 0.8600 |
| Pix2pix | 0.7561 | 0.4519 | 0.8068 | 0.8828 | **0.8320** | 0.8353 | **0.8176** | 0.8152 |
| MAGIC | **0.8302** | **0.9313** | **0.8158** | 0.9037 | 0.7938 | **0.9453** | 0.7975 | **0.9066** |

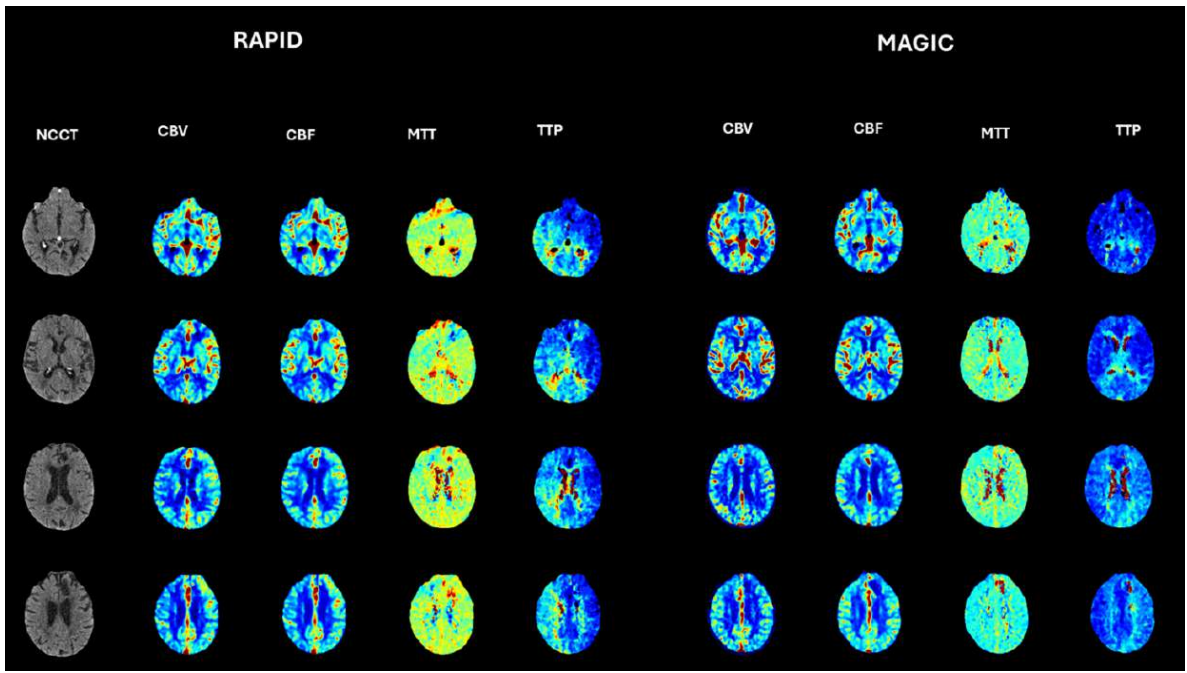

Figure S4: Example of small-vessel ischemic injury on non-contrast CT imaging. This figure demonstrates the substantial low density of the periventricular white matter, especially near the occipital horns, which is typically indicative of small vessel ischemic injury. Despite this, the perfusion parameters measured in the current MAGIC iteration do not reflect this condition, highlighting a limitation of the proposed method in detecting small-vessel occlusions based solely on non-contrast imaging.


\end{document}